\documentclass[preprint]{imsart}

\RequirePackage[OT1]{fontenc}
\RequirePackage{amsthm,amsmath,graphicx,multirow}
\RequirePackage[numbers]{natbib}
\RequirePackage[colorlinks,citecolor=blue,urlcolor=blue]{hyperref}
\usepackage[noend]{algpseudocode}
\algnewcommand\algorithmicto{\textbf{to}}
\algdef{SE}[FOR]{ForTo}{EndFor}[2]{%
  \algorithmicfor\ #1\ \algorithmicto\ #2\ \algorithmicdo
}{%
  \algorithmicend\ \algorithmicfor
}\usepackage{color}

\newcommand{\argmin}{\mathop{\rm arg~min}\limits}

\startlocaldefs
\numberwithin{equation}{section}
\theoremstyle{plain}
\newtheorem{thm}{Theorem}[section]
\newtheorem{prop}{Proposition}[section]
\newtheorem{lem}{Lemma}[section]
\newtheorem{rem}{Remark}[section]
\theoremstyle{remark}
\endlocaldefs

\newcommand{\eff}{\mathrm{eff}}

 \newcommand{\lp}{\left(}
 \newcommand{\rp}{\right)}
 \newcommand{\llp}{\left\{}
 \newcommand{\rrp}{\right\}}
 \newcommand{\lllp}{\left[}
 \newcommand{\rrrp}{\right]}
 \newcommand{\pd}{\partial}

 \newcommand{\mI}{\mathcal{I}}
\newcommand{\logit}{\mathrm{logit}}
\newcommand{\rmu}{\mathrm{u}}
\newcommand{\rmg}{\mathrm{g}}
\newcommand{\rmm}{\mathrm{m}}

\begin{document}

\begin{frontmatter}
\title{Semiparametric Optimal Estimation With Nonignorable Nonresponse Data}
\runtitle{Estimation With NMAR Data}

\begin{aug}
\author{\fnms{Kosuke} \snm{Morikawa}\thanksref{t1}\ead[label=e1]{morikawa@sigmath.es.osaka-u.ac.jp}}
\and
\author{\fnms{Jae Kwang} \snm{Kim}\thanksref{t2}\ead[label=e2]{jkim@iastate.edu}}

\thankstext{t1}{Supported by JSPS KAKENHI Grant 19K14592.}
\thankstext{t2}{Supported in part by U.S. NSF Grant MMS-1733572.}
\runauthor{K. Morikawa and J. K. Kim}

\affiliation{Osaka University\thanksmark{m1}, Iowa State University\thanksmark{m2}}

\address{Graduate School of Engineering Science\\
Osaka University\\
Toyonaka, Osaka 5608531\\
Japan\\
\printead{e1}}

\address{
Department of Statistics\\
Iowa State University\\
Ames, Iowa 50011\\
USA\\
\printead{e2}}
\end{aug}

\begin{abstract}
When the response mechanism is believed to be not missing at random (NMAR), a valid analysis requires stronger assumptions on the response mechanism than standard statistical methods would otherwise require. Semiparametric estimators have been developed under the model assumptions on the response mechanism. In this paper, a new statistical test is proposed to guarantee model identifiability without using instrumental variable assumption. Furthermore, we develop optimal semiparametric estimation for parameters  such as the population mean. Specifically, we propose two semiparametric optimal estimators that do not require any model assumptions other than the response mechanism. Asymptotic properties of the proposed estimators are discussed. An extensive simulation study is presented to compare with some existing methods. We present an application of our method using Korean Labor and Income Panel Survey data.
\end{abstract}



\end{frontmatter}

\section{Introduction}
\label{sec1}

Handling missing data often requires some assumptions about the response mechanism.
If the study variable does not affect the probability of the response, the response mechanism is called missing at random (MAR) \citep{rubin76}. If, on the other hand, the response probability of a study variable depends on that variable directly, the response mechanism is called not missing at random (NMAR) \citep{little02}. Under NMAR, the response probability cannot be verified using  the observed study variables only, therefore,  additional assumptions about the study variable are often required.
%An overview of the existing method Chapter of
%Model identification is challenging ( ) and model diagnostic tool is also limited.

 Let $r$ be the response indicator of the study variable $y$ with auxiliary variable $x$, where $r$ takes $1$ if $y$ is observed, and takes $0$ otherwise. In this paper, we consider a situation where the study variable $y$ is subject to missingness.  Ignorable nonresponse or MAR can be understood as the conditional independence of $r$ and $y$ given $x$, namely $r \perp y \mid x$, which is usually untestable. \citet{greenlees82} and \citet{diggle94} proposed a fully parametric approach to analyze nonignorable nonresponse data; their method requires two parametric models: (i) an outcome model, $[y \mid x]$; and (ii) a response model $[r\mid x, y]$. In practice, it is difficult to verify models (i) and (ii), because some of $Y$ are not observed. For the fully parametric approach, model identification and model misspecification can be a problem, and sensitivity analysis becomes necessary \citep{scharfstein99,rotnitzky01,verbeke01,tsiatis06}.
 \citet{sverchkov08} and \citet{riddles16} proposed a fully parametric approach that uses different model specifications based on (i) $[y \mid x, r=1]$, and (ii) $[r \mid x, y]$. Their approach is attractive because one can verify a model for $[y \mid x, r=1]$ from the  observed responses; however, because it is a fully parametric approach, it is still subject to model misspecification error.
 %Furthermore, model identification is also very challenging \citep{miao16}.

 Recently, several semiparametric approaches have been proposed for nonignorable nonresponses.
 %by which requires correct specification of the model in (i) only .
\citet{ma03} studied identification and parameter estimation for binary study variables. \citet{tang03} also considered model identification using an instrumental variable and proposed  a maximum pseudo likelihood estimator that does not require model specification of the response mechanism. \citet{dhau10} also used the same instrumental variable assumption and considered  a regression analysis using the nonparametric propensity score model. \citet{zhao15} extended the method of \citet{tang03} and relaxed the condition on the instrumental variable, which is called nonresponse instrumental variable \citep{wang14}. \citet{fitzmaurice05} and \citet{skrondal14} proposed protective estimators that do not require a model for the response mechanism, but the application of this approach is limited to situations in which $Y$ is binary.  In the meantime, \citet{kim11_2} proposed a semiparametric method for estimating $E(Y)$ using a semiparametric response model, but a validation sample is required in order to estimate the parameters in the response mechanism. \citet{tang14} used the method of empirical likelihood to
extend the method of \citet{kim11_2} to estimate more general parameters. In \citet{zhao17},
the method of \citet{qin02} was used to construct a $n^{1/2}$-consistent estimator without  a validation sample. \citet{morikawa17} used the kernel regression estimator to remove the parametric model assumption on model (i) $[y \mid x, r=1]$. \citet{chang08} and \citet{wang14} considered a generalized method of moments (GMM) estimator that uses the response model assumption only, but their method is generally lacking in efficiency. Recently, \citet{shao16} proposed a semiparametric inverse propensity weighting method using the nonresponse instrumental variable (NIV) assumption of \citet{wang14}.  However, the above papers do not address efficiency of their semiparametric estimation methods. Furthermore, the NIV assumption is difficult to verify from the sample. Developing an optimal semiparametric estimator and a test procedure for model idenitification under NMAR are important research topics in missing data analysis.
%\citet{miao16} considered doubly robust estimators under nonignorable nonresponse. Their estimator can achieve consistency and asymptotic normality if the conditional odds ratio model is true and if either the baseline outcome density model or the baseline response model is true. In other words, the doubly robust estimators require that two out of the three aforementioned models be correctly specified.

In this paper we use a parametric model for $[r \mid x, y]$ and  a fully nonparametric model for $[y\mid x, r=1]$ to form a semiparametric model and develop a nonparametric test procedure for model identification of the semiparametric model. After that, we  construct optimal estimators for parameters both related to the response mechanism and for the parameter of interest such as population mean. Efficiency under this setup has already been discussed by \citet{rotnitzky97} and \citet{robins99}. However, their estimator requires many working models to achieve the semiparametric efficiency bound. Misspecification of the working models may lead to loss of efficiency. See the simulation study in \S 6 and real data analysis in \S 7 for comparison with the method of \citet{rotnitzky97}.

Therefore, we consider an alternative approach and propose two semiparametric estimators that attain the semiparametric lower bound \citep{bickel98} (1) with a working model assumption or (2) without requiring working model assumptions. The first estimator is an adaptive estimator using a working model for $[y\mid x, r=1]$. If the working model is correct, the first estimator attains the lower bound. The second one is based on the nonparametric regression model which does not require any additional assumptions, but it still attains the lower bound. All technical details are given in Appendix \ref{sup.B}.

\section{Basic setup}
\label{sec2}

Let $(z_i, r_i),\;i=1,\ldots, n$ be $n$ realizations from a joint distribution $[z, r]$, where $z=(x^\mathrm{T}, y)^\mathrm{T}$, $x$ is a $d$-dimensional covariate vector, $y$ is a response variable, and $r$ is a response indicator of $y$, i.e., it takes $1$ if $y$ is observed, and takes 0 otherwise.  Also, let $G_r(z)$ be the observed data when the response indicator is $r$, i.e., $G_1(z)=z$ and $G_0(z)=x$. Suppose that the response model is $\pi(z;\phi)$ with a $q$-dimensional parameter $\phi\in\Phi$. Let $\theta\in\Theta$ be a parameter satisfying $E\{U(Z; \theta)\}=0$, where $U$ is a known function of $z$. For example, if we are interested in $E(Y)$, then $U(z;\theta)=y-\theta$, and in regression coefficients $E(Y\mid x)=\mu(x; \theta)$, then $U(z;\theta)=a(x)\{y-\mu(x;\theta)\}$, where $a(\cdot)$ is any linearly independent function of $x$ having same dimension as $\theta$. In this paper, we consider semiparametric estimation of $(\phi, \theta)$ from partial observations. In particular, we propose the efficient estimator among the regular asymptotically linear estimators \citep{bickel98, tsiatis06} without relying on the correctness of $U$ function and propose two adaptive estimators.

For model identification for a response model, \citet{miao16} gives a sufficient condition when the outcome models are normal or normal mixture. However, the normality assumption cannot be checked directly from observed data. In the meantime, \citet{wang14} developed a theory for identification by assuming that there exists a NIV $x_2$ in the covariate vector $x=(x_1^\mathrm{T}, x_2^\mathrm{T})^\mathrm{T}$such that $x_2$ is independent of $r$, given $x_1$ and $y$. When $x$ is the single variable, $x$ itself is the NIV. Although the existence of such a NIV is a sufficient condition, it is hard to verify it from the observed data. Therefore, both identification conditions are not testable with observed data. In \S \ref{sec3}, we propose an alternative condition for the model identification by assuming a restriction on [$y\mid x, r=1$], not only on the response mechanism, and develop a test procedure for model identification.

Classical approaches for analyzing nonignorable nonresponse data are based on correct specification for $[y\mid x]$ as well as the response mechanism \citep{greenlees82}. This requirement can be challenging because the specification cannot be verified under nonignorable nonresponse \citep{molenberghs08}. \citet{chang08} proposed a semiparametric estimator for $\phi$ based on the following estimating equation:
%does not require any additional assumptions:
\begin{align}
\sum_{i=1}^n\Gamma(x_i, y_i, r_i; \phi)=\sum_{i=1}^n\llp 1-\frac{r_i}{\pi(z_i;\phi)}\rrp g(x_i; \phi)=0, \label{ck}
\end{align}
where $g=\{g_1(x), g_2(x), \ldots g_q(x)\}^\mathrm{T}$, which can be called calibration function, is a function of $x$ whose elements are linearly independent; $q$ is the dimension of $\phi$. Note that although this estimator satisfies consistency and asymptotic normality under certain regularity conditions, its efficiency is not guaranteed. %If $q=d+1$, where $d$ is the dimension of $x$, a typical choice for $g$ is $g(x)=(1, x^\mathrm{T})^\mathrm{T}$. However, if $q>d+1$, it is  not clear how to determine the best $g$ in terms of minimizing the asymptotic variance. \citet{chang08} also discussed computation using generalized method of moments (GMM).

Recently, \citet{riddles16} proposed an efficient estimator that uses a parametric model for $[y\mid x, r=1]$. Using the mean score theorem \citep{louis82}, the maximum likelihood estimator can be obtained by solving
\begin{align*}
\sum_{i=1}^n  \left[  r_i s_1(z_i; \phi) + (1-r_i)E_0\{s_0 (Z; \phi) \mid x_i  \}  \right] =0,
\end{align*}
where $s_r(z; \phi)$ is the score function of $\phi$, that is,
\begin{align}
s_r(z; \phi)=\frac{\{r-\pi(z; \phi)\}\dot{\pi}(z; \phi)}{\pi(z; \phi)\{1-\pi(z; \phi)\}}, \label{sr}
\end{align}
$\dot{\pi}(z; \phi)=\pd \pi(z; \phi)/\pd\phi$, and $E_0(\cdot\mid x)$ is the conditional expectation conditional on $x$ and $r=0$. To compute $E_0(\cdot\mid x)$,  under Bayes' formula, \citet{riddles16} proposed using
\begin{align}
\sum_{i=1}^n\lllp r_i s_1(z_i; \phi) + (1-r_i)\frac{E_1\{O(Z; \phi)s_0(Z; \phi)\mid x_i\}}{E_1\{O(Z; \phi)\mid x_i\}}\rrrp=0, \label{2.3}
\end{align}
where $O(z; \phi)=\{1-\pi(z; \phi)\}/\pi(z; \phi)$, and $E_1(\cdot\mid x)$ is the conditional expectation on $y$ given $x$ and $r=1$. The conditional expectation is computed by assuming a parametric model $f_1(y\mid x;\gamma)=f(y\mid x, r=1; \gamma)$. This may increase the efficiency, however, misspecification of the $f_1$ model  could cause the solution $\hat{\phi}$ to be inconsistent. \citet{morikawa17} proposed a semiparametric method using a nonparameteric estimator of $f_1$, assuming that the semiparametric model is identified. We now give more rigorous treatments of the model identification of the semiparametric model.

\section{Identification}
\label{sec3}

We consider a new identification condition for estimation of the response model with observed data.  Our idea is to define the target parameter $\phi_0$ as a unique solution to
\begin{align}
	E\llp \Gamma(Z, R; \phi)\mid X\rrp =0 \quad \mathrm{a.s.}, \label{iden.eq}
\end{align}
where $\Gamma$ is defined in \eqref{ck}, though natural definition of the parameter might be through either (i) $E\{\Gamma (Z, R; \phi)\mid Z\}=0$ or (ii) $E\{\Gamma (Z, R; \phi)\}=0$. Note that providing a sufficient condition for the parameter defined in (ii) is the strongest (and  in (i) is the weakest) since $E\{\Gamma(Z, R; \phi)\}=E[E\{\Gamma(Z, R; \phi)\mid Z\}]$ and $E\{\Gamma(Z, R; \phi)\mid X\}=E[E\{\Gamma(Z, R; \phi)\mid Z\}\mid X]$ hold. This implies a sufficient condition for the parameter \eqref{iden.eq} does not necessarily guarantee the model identification of (ii), which is the probability limit of the estimating equation \eqref{ck}. However, it is rare in practice that the model \eqref{iden.eq} is identifiable, but the model (ii) is not. Also even if we face such a problem, it can be solved by constructing an objective function with the integrated regression function and additional minor conditions \citep[see][Assumptions 1--3]{dominguez04}.  For the above reasons, we focus on providing a sufficient condition of the model identification for the parameter defined in \eqref{iden.eq}.

%Section3.1st
\subsection{Identification condition with $f_1$ model}
Let  $O(z;\phi)=1/\pi(z;\phi)-1$ be the odds function of the response model, $E_1(\cdot\mid x)$ be the operator for the true conditional expectation given $x$ and $r=1$. 
A new identification condition for the semiparametric model is given in the following theorem.

\begin{thm}
\label{th.identification}
The identification condition for a parameter \eqref{iden.eq} holds under the following conditions.
\begin{enumerate}
\renewcommand{\theenumi}{(I\arabic{enumi})}
	\item \label{I1} $E_1\{O(Z;\phi)\;|\;x\}$ exists and is bounded almost surely;
	\item \label{I2}  The weight function $g$ in \eqref{ck} satisfies $P\lp \inf_{\phi\in\Phi}|g(X; \phi)|>0\rp > 0$, and  elements of $g(x; \phi)$ are linearly independent functions with respect to $x$ for all $\phi$;
	\item \label{I3}  $E_1\{O(Z;\phi)\mid x\}=E_1\{O(Z;\phi')\mid x\}$ a.s. implies $\phi=\phi'$.
\end{enumerate}
\end{thm}

By the condition \ref{I1}, the response model is almost limited to the logistic regression models. For example, let $\Psi$ be the cumulative function of the standard normal distribution. Assume that $\pi(y)=\Psi(y)$ (probit model) and the density of $f_1$ is the standard normal distribution, then 
\begin{align*}
E_1\{O(Y)\mid x\}&=E_1\llp \frac{1}{\pi(Y)} -1\rrp\\
&=  \int_{-\infty}^{\infty} \frac{\pd }{\pd y}  \log \Psi(y) dy -1,
\end{align*}
and $E_1\{O(Y)\mid x\}$ does not exist even for this simple probit model. Nevertheless, Theorem \ref{th.identification}
 is practically useful because the performance with the probit  and logistic model is very similar, and thus, misspecification of the response model is not a serious problem in practice (see \S 7 for the performance with misspecified response models). Condition \ref{I2} is required to avoid $g$ becomes identically zero. 

The key condition is \ref{I3}, which implies that we should check the identification of $E_1\{O(Z;\phi)\mid x\}$. Checking the identification of $E_1\{O(Z;\phi)\mid x\}$ is relatively easy and feasible with observed data. For example, if the response mechanism is specified as $\pi(z; \phi)=1/\{1+\exp(\phi_{\mathrm{x}0}+\phi_{\mathrm{x}1}x+\phi_{\mathrm{y}}y)\}$, where $\phi=(\phi_{\mathrm{x}0}, \phi_{\mathrm{x}1}, \phi_{\mathrm{y}})^\mathrm{T}$.
Then, $E_1\{O(Z;\phi)\;|\;x\}$ is written as
\begin{align}
E_1\{O(Z;\phi)\;|\;x\}=\exp\{\phi_{\mathrm{x}0}+\phi_{\mathrm{x}1}x+K_{\phi_{\mathrm{y}}}(x)\}, \label{id.cond}
\end{align}
where $K_{\phi_{\mathrm{y}}}(x)=\log E_1\{\exp(\phi_{\mathrm{y}}Y)\mid x\}$ is the cumulant-generating function of $[y\mid x, r =1]$. Therefore, we have only to check whether $K_{\phi_{\mathrm{y}}}(x)$ is linear with respect to  $x$ or not. If $f_1$ is a parametric model, the model identification for $\phi$ is easy to check. For example,  if $[y\mid x, r=1]$ belongs to an exponential family with the density function
\begin{align*}
f_1(y\mid x; \tau, \psi)=\exp\left[\frac{y\tau(x)- b\{\tau(x)\}}{\psi}+c(y,\psi)\right],
\end{align*}
where  $\psi$ is the dispersion parameter and $\tau$, $b$, $c$ are known functions, then the cumulant-generating function reduces to $K_{\phi_{\mathrm{y}}}(x)=\{b(\phi_{\mathrm{y}}\psi+\tau(x))-b(\tau(x))\}/\psi$, from which we can verify the model identification. For example, for model identification, $b$ is allowed to be any polynomial function except for the 1st- and 2nd-order function of $x$ such as log-function (e.g. Gamma distribution), exponential-function (e.g. Poisson distribution), etc. However, when $b$ is a 2nd-order polynomial function, for example, $b(\tau)=\tau^2/2$, which means $f_1$ follows normal distribution, then $K_{\phi_{\mathrm{y}}}(x)=\tau(x)\phi_{\mathrm{y}} + \phi_{\mathrm{y}}^2\psi^2/2$. Also,  we obtain
$$E_1\{O(Z;\phi)\;|\;x\}=\exp\{\phi_{\mathrm{x}0}+\phi_{\mathrm{x}1}x+ \tau(x)\phi_{\mathrm{y}} + \phi_{\mathrm{y}}^2\psi^2/2\}.$$
Thus, by Theorem \ref{th.identification}, $\phi$ is identifiable unless the mean structure $\tau(x)$ is linear since there are three parameters with two equations. If $\tau(x)$ is linear, we may use a transformation approach which is introduced in \S \ref{sec3_3}.

On the other hand, checking the model identifiability with a nonprametric $f_1(y\mid x)$ model is still challenging, because there is no way to estimate the cumulative function $K_{\phi_{\mathrm{y}}}(x)$ nonparametrically for every $\phi_{\mathrm{y}}$ up to our knowledge.  Therefore, we propose a test statistic to test a reasonable necessary condition for the identification condition.

%Section3.2
\subsection{Nonparametric test statistics}
\label{s3.2}

In view of \eqref{id.cond}, the model is unidentifiable when the cumulative function is linear with respect to $x$ for all $\phi_{\mathrm{y}}$, i.e., the null hypothesis $H_0$:
$K_{\phi_{\mathrm{y}}}(x)=c_1(\phi_{\mathrm{y}})+c_2(\phi_{\mathrm{y}})x$ holds, where $c_1(\phi_{\mathrm{y}})$ and $c_2(\phi_{\mathrm{y}})$ are functions of $\phi_\mathrm{y}$. If $c_1$ and $c_2$ can be infinitely differentiable at $\phi_{\mathrm{y}}=0$, we have $K^{(\ell)}_{0}(x) = c^{(\ell)}_1(0)+c^{(\ell)}_2(0)x$ for all $\ell=1,2,\ldots$, where the superscript stands for the $\ell$-th partial derivative with respect to $\phi_{\mathrm{y}}$. Because the cumulant-generating function is expanded as
$$K_{\phi_{\mathrm{y}}}(x)=\sum_{\ell=0}^\infty\frac{\phi_{\mathrm{y}}^\ell}{\ell!}K^{(\ell)}_{0}(x),$$%=\sum_{\ell=0}^\infty\frac{\phi_{\mathrm{y}}^\ell}{\ell!}\{c^{(\ell)}_1(0)+c^{(\ell)}_2(0)x\}.$$
the linearity of the cumulant-generating function can be checked by that of $K^{(\ell)}_{0}(x)$ for all $\ell$. Based on this idea, we obtain an alternative null hypothesis $H^{(L)}_0: K^{(\ell)}_{0}(x) = c^{(\ell)}_1(0)+c^{(\ell)}_2(0)x,~ \ell=1, \ldots, L,$ for a positive integer $L$ or $L=\infty$. When $L=1$, this corresponds to a goodness-of-fit test of a simple linear regression with a normal distribution in $f_1$. %Thus, the test with $L>1$ is a generalization of the $f_1$ model to more general models. 
Although $L=1$ is just a necessary condition for nonparametric models,  in many cases, it is enough to guarantee the model identification.

Let a general data-generating process be $y=\mu(x)+\varepsilon(x)$, where $\mu(x)$ is the conditional expectation of $y$ given $x$, and $\varepsilon(x)$ is the conditonal mean-zero error. Consider a class of error functions $\mathcal{E}$: for $\varepsilon\in\mathcal{E}$, $\varepsilon(x)=\sum_{j=0}^\infty \xi_j e_j(x)$, where $\xi_j\;(j\geq 0)$ are mean-zero random variables which are independent of $x$, and $e_j\;(j\geq 0)$ are any measurable functions of $x$ satisfying $E[\{\sum_{j=0}^\infty|\xi_je_j(X)|\}^k]<\infty$ for any positive integer $k$, and $e_k\neq e_l$ for $k\neq l$. This class of error functions include many functions with mean-zero conditional expectation such as the infinite normal mixture distribution.  Under this setup, we can show the following proposition. 

\begin{prop}
	\label{prop3.1}
	Suppose that $\varepsilon\in\mathcal{E}$, then $H^{(\infty)}$ implies $H^{(1)}$.
\end{prop}

For the above reasons, we test a data-generation structure 
\begin{align}
y=\mu(x)+\varepsilon, \label{indep_model}
\end{align}
where $\mu(x)$ is a linear function of $x$, and $\varepsilon$ is a mean-zero random variable and independent of $x$, and consider a test to check the goodness-of-fit of the linear model. It is desirable that the statistical test enjoys two properties: (i) dimension free for $x$; (ii) no parametric assumption on $\varepsilon$. The first property is practically useful because classical nonparametric tests such as \citet{eubank92} suffer from curse of dimensionality. The second property can avoid subjectivity imposing some parametric assumption on the error variable. Recently, some nonparametric methods to check a goodness-of-fit have been proposed with  Hilbert-Schmidt independence criterion (HSIC)\citep{gretton05, gretton08, hidalgo18, sen14} and  mutual information  \citep{berrett19}. In this paper, we utilize an idea of  HSIC proposed by \citep{gretton05, gretton08}. With HSIC, \citet{sen14} and \citet{hidalgo18} proposed a test statistics to check goodness-of-fit of a (parametric/nonparametric) model, which has the two desirable properties.  Their idea is based on the fact that independence of $X$ and $\varepsilon$ implies correctness of the mean function $\mu(x)$ because $\epsilon$ is independent of $x$. The HSIC can be used to check the independence.

Let $\mathcal{F}$ be a reproducing kernel Hilbert space (RKHS) on a domain $\mathcal{X}$ with  a positive-definite function $k: \mathcal{X}\times \mathcal{X}\to \mathcal{R}$. The Hilbert space $\mathcal{F}$ has inner product $\langle \cdot, \cdot \rangle$ satisfying a property called reproducing property $\langle f, k(x, \cdot)\rangle=f(x)\;(f\in \mathcal{F}, x\in \mathcal{X})$. The kernel mean on the RKHS is defined by $E[k(\cdot, X)]=\int k(\cdot, x)dP(x)$ where $P$ is the probability measure of a random variable $X$. Then, a kernel $k$ is called characteristic if the kernel mean determines the probability measure $P$ uniquely. For example, \citet{fukumizu04} showed the gaussian kernel $k(x, \tilde{x})=\exp(-\sigma^{-1}\|x-\tilde{x}\|)$ is characteristic, where $\sigma$ is a tuning parameter and median is often used as a heuristic estimate of $\sigma$. Next, define the HSIC. Let $\mathcal{G}$ be another characteristic RKHS with kernel $l$. Then, define HSIC of between two random variables $X$ and $Y$, $ M_{XY}$,  by
 \begin{align*}
M_{XY}&=E\{k(X, \tilde{X})l(Y, \tilde{Y})\}+E\{k(X, \tilde{X})\}E\{l(Y, \tilde{Y})\}\\
& \quad -2E\lllp E\{k(X, \tilde{X})\mid X\}E\{l(Y, \tilde{Y})\mid Y\}\rrrp,
 \end{align*}
where $(\tilde{X}, \tilde{Y})$ is independent copy of $(X, Y)$. \citet{gretton05} shows that if the product kernel $kl$ is characteristic, $M_{XY}=0$ implies independence between $X$ and $Y$. By checking $M_{X\varepsilon}=0$ in the model \eqref{indep_model}, goodness-of-fit of a mean function $\mu(x)$ can be tested with observed data. The HSIC $M_{XY}$ is estimated with a $V$-statistics based estimator, $\hat{M}_{XY}=n^{-2}\mathrm{tr}(KHLH)$, where $K_{ij}=k(X_i, X_j)$, $L_{ij}=l(Y_i, Y_j)$, $H=I_n-n^{-1}1_n1_n^\top$, $I_n$ is the $n\times n$ identity matrix, and $1_n$ is the $n\times 1$ vector of ones. Unlike $\hat{M}_{XY}$, it is hard to derive the asymptotic distribution of $\hat{M}_{X\varepsilon}$ under the null hypothesis because $\mu$ in \eqref{indep_model} is replaced with an estimated mean function, and the limiting distribution becomes more complicated. However, the bootstrap method is applicable to estimate the distribution as follows \citep{sen14}. In the algorithm, let $x=(x_1, \ldots, x_{n1})^\top$ and $y=(y_1, \ldots, y_{n1})$ be observed covariate variables and response variables, and $\mu(x_i; c)=c_1+c_2 x_i$, where $c=(c_1, c_2)^\top$ be a linear model under the null hypothesis. To make the algorithm simple and clear,  a vector is used instead of use of each element, that is, $\mu(x; c)$ implies the vector $(\mu(x_1; c), \ldots, \mu(x_{n_1}; c))^\top$.

    \begin{algorithmic}[1]
        \Procedure{Bootstrap Method for $\hat{M}^{(b)}_{X\varepsilon}\;(b=1, \ldots, B)$}{}
        \State{$\hat{c} \leftarrow \argmin_c \sum_{i=1}^{n_1}\{y_i-\mu(x_i; c)\}^2$}
        \State{$\hat{\varepsilon}\leftarrow y-\mu(x; \hat{c})$}
        \ForTo{$b = 1$}{$B$}
            \State{$x^{(b)}\leftarrow$ bootstrap sample from observed data $x$}
            \State{$\varepsilon^{(b)}\leftarrow$ bootstrap sample from $\hat{\varepsilon}$}
            \State{$y^{(b)}\leftarrow\mu(x^{(b)}; \hat{c})+\hat{\varepsilon}^{(b)}$}
            \State{$\hat{c}^{(b)}\leftarrow\argmin_c \sum_{i=1}^{n_1}\{y^{(b)}_i-\mu(x^{(b)}_i; c)\}^2$} 
            \State{$\hat{\varepsilon}^{(b)}\leftarrow y^{(b)}-\mu(x^{(b)}; \hat{c}^{(b)})$}
            \State{$(K^{(b)})_{ij}\leftarrow k(x^{(b)}_i, x^{(b)}_j)$;~ $(E^{(b)})_{ij}=k(\hat{\varepsilon}^{(b)}_i, \hat{\varepsilon}^{(b)}_j) \quad \mathrm{for~all~}i,j$} 
%            \State{$H=I_n-n^{-1}1_n1_n^\top$}
            \State{$\hat{M}^{(b)}_{X\varepsilon}=n^{-2}_1\mathrm{tr}(K^{(b)}HE^{(b)}H)$}
             \EndFor
        \EndProcedure
    \end{algorithmic}

Then, we have the following asymptotic result.
\begin{prop}
\label{test.identification}
Let $\hat{M}_{X\varepsilon}^*$ be the bootstrap test statistics for $M_{X\varepsilon}$. Suppose that the kernels $k$ and $l$, which prescribe RKHS of the random variables $X$ and $Y$, and the mean function $\mu(x)$ satisfies Condition 1, 2, and 5 in \citet{sen14}. Then, under the null hypothesis $H^{(1)}_0$, the asymptotic distribution of $n_1\hat{M}^*_{X\varepsilon}$ is the same as that of $n_1M_{X\varepsilon}$.
\end{prop}

%Section3.3
\subsection{Doubly-normalized exponential transformation}
\label{sec3_3}

When the null hypothesis is not rejected, an instrumental variable is required to make the estimator consistent. However, selecting the instrumental variable is very difficult even if it exists.  Because the problem comes from using the same covariate between the response model and mean function, we can make an identifiable model artificially by transforming covariate variable $x$ in a response model to a nonlinear variable $\mathcal{T}(x)$ such as $\exp(x)$ and $x^2$, at the sacrifice of consistency. Although there are many choices of such functions, it would be desirable that the transformation enjoys three properties: (i) ``nonlinearity" can be adjusted through a tuning parameter $a$ such that  $\lim_{a\to 0}\mathcal{T}_a(x)=x$; (ii) the value $a$ does not depend on range/scale of $x$; (iii) range of $\mathcal{T}_a(x)$ is same as that of $x$.
The first condition is necessary to adjust ``nonlinearity": small $a$-value holds the original data structure, and large $a$-value breaks the structure, but provides stronger identification. For example, one may come up with a transformation $\mathcal{T}_a(x)=\log \{\exp(a+x)\}$. However, nonlinearity of such a transformation may heavily depend on both $a$ and range/scale of $x$ so that it is necessary to find an appropriate value $a$ (which is close to 0) for every covariate or dataset, hence, the second condition is required. The third condition is requisite to retain the value of response probability to some extent. Considerably large (small) value of $\mathcal{T}_a(x)$ may damage the bounded condition $\pi(\mathcal{T}_a(x), y)>0$, which is often assumed in this field.  

\begin{figure}[h!]
	 \begin{center}
	\includegraphics[width=110mm]{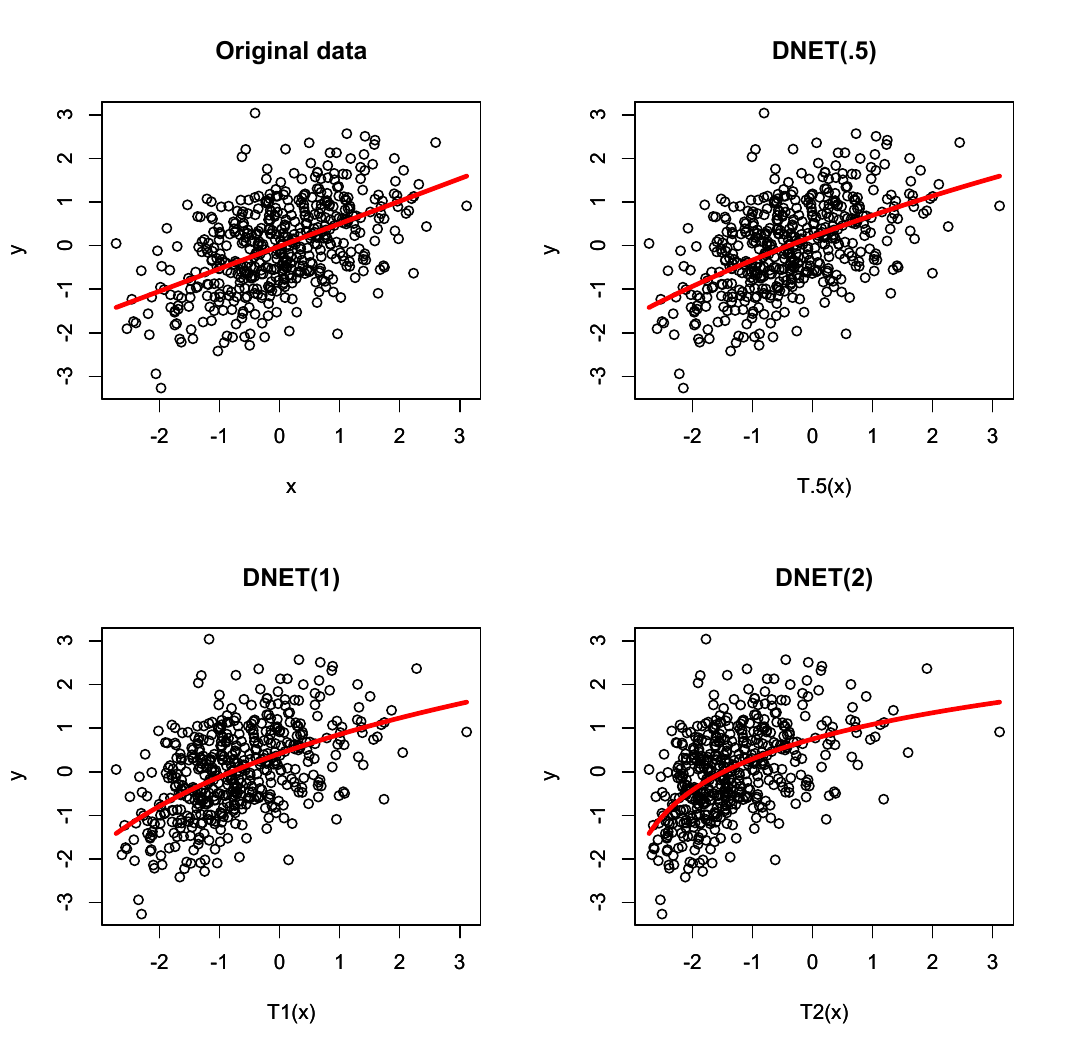}
	  \end{center}
	\caption{Illustration of DNET: each top left, top right, bottom left, and bottom right shows the scatterplot of $\mathcal{T}_{a}(x)$ v.s. $y$, for $a=0$(original), $0.5, 1, 2$, respectively. The red curve is conditional mean function of $y$ given $\mathcal{T}_a(x)$.}
	\label{fig:1}
\end{figure}

We propose a simple nonlinear transformation having three desirable properties called doubly-normalized exponential transformation (DNET). Let $\mathcal{S}_a(x)$ be a normalized exponential transformation 
$\mathcal{S}_a(x)=\{\mathrm{Var}(aX)\}^{-1/2}\{\exp(a x)-E(\exp(aX))\}$.
By letting $a\to0$, we obtain
\begin{align*}
\lim_{a\to 0}\mathcal{S}_a(x)&=\lim_{a\to 0}\frac{a^{-1}\{\exp(a x)-1\}+a^{-1}\{1-E(\exp(aX))\}}{\{\mathrm{Var}(X)\}^{1/2}}\\
&=\frac{x-E(X)}{\{\mathrm{Var}(X)\}^{1/2}}.
\end{align*}
This indicates that the normalized exponential transformation $\mathcal{S}_a$ after data normalization is an identity map as $a\to 0$, i.e., with $\mathcal{Z}: x\mapsto \{\mathrm{Var}(X)\}^{-1/2}(x-E(X))$, a map $\mathcal{S}_a\circ\mathcal{Z}$ becomes identity as $a\to0$. Finally, after some minor modification to satisfy the third condition above, we have our proposed transformation method:
    \begin{algorithmic}[1]
        \Procedure{Compute DNET}{$a$}
            \State{$z\leftarrow\{\mathrm{var}(x)\}^{-1/2}(x-\mathrm{mean}(x))$}
            \State{$s\leftarrow\{\mathrm{var}(az/5)\}^{-1/2}\{\exp(a z/5)-\mathrm{mean}(\exp(a z/5))\}$}
        \State $r_x\leftarrow\max(x)-\min(x);~ r_s\leftarrow\max(s)-\min(s)$
            \State{$\mathcal{T}_a(x)\leftarrow\min(x)+\{s-\min(s)\}\times r_x/r_s$}
        \EndProcedure
    \end{algorithmic}
In the algorithm, each mean, var, max, and min is  sample mean, variance, maximum and minimum value of $x=(x_1, \ldots, x_n)$. Obtained $\mathcal{T}_a(x_i)\;(i=1, \ldots, n)$ is the proposed nonlinear transformation. The reason divided by 5 is just for scale adjustment. We call the transformation with $a$-value 0.5(weak), 1(moderate), and 2(strong) nonlinearity.  In Figure \ref{fig:1}, we illustrate the scatterplot of $\mathcal{T}_{a}(x_i)$ versus $y_i$, for $a=0$(original), 0.5, 1, 2, where $(x_i, y_i)\;(i=1, \ldots, 500)$ are independently generated from a bivariate normal distribution with both mean 0, variance 1, and correlation 0.5. It can be seen that the transformation enjoys the three desirable properties.

%\begin{rem}
%{
%Replacing $\exp(aX/3)$ with $\log\{1/a+x-\min(x)\}$ also leads to a nonlinear transformation possessing the three desirable properties. It works like an inverse function of DNET. In this paper, we only focus on DNET.
%}
%\end{rem}

%Section4
\section{Efficiency Bound}

In this section, we provide an optimal influence function for the true parameter $(\phi_0^\mathrm{T}, \theta_0)^\mathrm{T}$ that is the most efficient among all regular and asymptotically linear estimators, but that does not depend on the correctness of the $U$-function, i.e., we put a  constraint that the nuisance tangent space of $\theta$ and $\phi$ are orthogonal. For example, \citet{rotnitzky97} derived the semiparametric efficiency bound for regression parameters, which prescribe the first moment of the distribution of $[y\mid x]$. However, their adaptive estimators require many working models, and misspecification of either a regression model or a response model leads to a biased estimator, but in most cases, we do not expect the regression model is true and assume a simple function such as a linear regression model. In this section, we first provide the efficiency bound under the response model only, without relying on the information of $U$-function because the most difficult task in nonignorable nonresponse missing data analysis is to obtain a consistent estimator of the response model. Optimal estimators achieving this lower bound will be considered in the next section.
%\textcolor{red}{(I am not sure whether this is accurate. - JK ) }\textcolor{blue}{(This is correct. -KM)}

In the following discussion, we abbreviate the parameter value or random variable, for example, $\pi(z;\phi_0)=\pi(z)=\pi(\phi_0)$, unless this would lead to ambiguity.

\begin{lem}
\label{Lem.eff}
Let  $S_\eff=(S^\mathrm{T}_1, S_2)^\mathrm{T}$, where
$S_1= S_1(R, G_R(Z))$ and $S_2= S_2(R, G_R(Z))$ be  defined as
\begin{align}
S_1(R, G_R(Z); \phi)&=\llp 1-\frac{R}{\pi(Z;\phi)}\rrp g^{\star}(X; \phi_0), \label{3.1.1}\\
S_2(R, G_R(Z);\phi, \theta)&=\frac{R}{\pi(Z;\phi)} U(Z; \theta)+\llp 1-\frac{R}{\pi(Z;\phi)}\rrp U^{\star}(X; \phi_0, \theta)\},\label{3.1.2}
\end{align}
$g^{\star}(x; \phi_0)=E^{\star}\{s_0(Z;\phi_0)\mid x; \phi_0\}$, $U^{\star}(x; \phi_0, \theta)=E^{\star}\{U(Z;\theta)\mid x; \phi_0\}$, and
\begin{align}
E^{\star}\{g(Z)\mid x; \phi_0\}=\frac{E\{O(Z; \phi_0)g(Z)\mid x\}}{E\{O(Z; \phi_0)\mid x\}} \label{estar}
\end{align}
with  $O(z; \phi_0)=\{1-\pi(z; \phi_0)\}/\pi(z; \phi_0)$.
Then, the efficient influence function is $\varphi_\eff=H^{-1} S_\eff$, where
$H=E(S_\eff^{\otimes 2})=E\llp \pd S_\eff(\phi_0, \theta_0)/\pd (\phi^\mathrm{T}, \theta)^\mathrm{T}\rrp$
and $B^{\otimes 2}=B B^\mathrm{T}$. Therefore, the semiparametric efficiency bound is given by $\{E(S_\eff^{\otimes 2})\}^{-1}$.
\end{lem}

This lemma implies that if we can compute $E^{\star}(\cdot \mid x)$ then estimating functions \eqref{3.1.1} and \eqref{3.1.2} will provide an optimal estimator. The optimal estimator is the solution to
\begin{align}
\sum_{i=1}^n S_{\eff, i}(\phi, \theta)=\sum_{i=1}^n\{S^\mathrm{T}_1( r_i, G_{r_i}(z_i);\phi), S_2(r_i, G_{r_i}(z_i); \phi,\theta)\}^\mathrm{T}=0. \label{3.1.3}
\end{align}

The equation based on $S_1(\phi)$ in \eqref{3.1.1} gives an optimal estimator for $\phi$, say $\hat{\phi}$. Then, by using $\hat{\phi}$, $S_2(\hat{\phi}, \theta)$ in \eqref{3.1.2} can provide an optimal estimator for $\theta$. However, the expectation $E^\star(\cdot \mid x)$ and the parameter $\phi_0$ are unknown and need to be estimated. Also, to compute the conditional expectation, we may need to correctly specify the distribution of $[y\mid x]$, which is subjective and unverifiable, as is stated in \S \ref{sec1}. In the next section, two adaptive estimators are proposed to work around the problem and to attain the lower bound derived in Lemma \ref{Lem.eff}.

\begin{rem}
Equation \eqref{3.1.1} can be viewed as a special case of the estimator of \citet{chang08} defined in \eqref{ck}. Thus, the optimal $g$ function in \eqref{ck} for the \citet{chang08} method  is given by $g^{\star}(x, \phi_0)$ in (\ref{3.1.1}) although $\phi_0$ is unknown. %\citet{qin02} also derived an estimator that is effective for nonignorable nonresponse data. However, it turns out that their estimator is exactly the same as that of \citet{chang08} when $q=d+1$, where $q$ and $d$ are the dimensions of $\phi$ and $x$, respectively. The proof is shown in \S S3 in the Supplementary Material. Also
One might think that the efficiency can be improved with a larger dimension of $g$ because the above two methods can handle over-identified models with $q>d+1$. However, according to Lemma \ref{Lem.eff},  there is no need to use more $g$ functions and it is enough to consider only $g^{\star}(x, \phi_0)$ (i.e., $q=d+1$) as the calibration function.
\end{rem}

\begin{rem}
The optimal score function in \eqref{3.1.1} can be derived differently as follows. Consider the class of estimating equations in \eqref{ck} indexed by $g$. For given $g$, the asymptotic variance of the solution $\hat{\phi}_g$ to \eqref{ck} can be written as
$$ V ( \hat{\phi}_g) = \frac{1}{n} A_g^{-1} B_g A_g^{-1} $$
where
\begin{eqnarray*}
A_g  & = & E \left[   E\{ O (Z; \phi_0) \cdot s_0 (Z; \phi_0) \mid X \} g(X; \phi_0)^\mathrm{T} \right] \\
B_g &=& E\left[E\{ O (Z; \phi_0) \mid X \}  g(X; \phi_0) g(X; \phi_0)^\mathrm{T} \right] .
\end{eqnarray*}
Using Cauchy-Schwarz inequality, the asymptotic variance is minimized at $g^{\star}(x; \phi_0)=E^{\star}\{s_0(Z;\phi_0)\mid x; \phi_0\}$. Similarly, we can obtain the optimal estimating function in \eqref{3.1.2} by considering a class of estimating equations of the form
\begin{equation}
\sum_{i=1}^n \left[ \frac{ r_i}{\pi (z_i; \phi) } U(z_i ; \theta) +\left\{ 1- \frac{r_i}{\pi(z_i; \phi) } \right\} h(z_i) \right] =0 ,
\label{cc}
\end{equation}
indexed by $h$. The asymptotic variance of the solution to (\ref{cc}) is minimized at
$h = E^{\star} \{ U( Z; \theta) \mid x; \phi_0 \}$.
\end{rem}

\begin{rem}
In our estimation steps, $\phi$ and $\theta$ are separately estimated. Thus, it follows from the identifiability of $\phi$ that $\theta$ is also identifiable. This is because, under assumptions \ref{I1}--\ref{I3}, $\phi$ is identifiable, thus, the identification problem of $\theta$ reduces to that of the probability limit of \eqref{3.1.3} or expectation of \eqref{3.1.2}, i.e., $E\{U(Z;\theta)\}$. 
\end{rem}

%\begin{rem}
%If $\theta$ is a regression coefficient, because it is related to the distribution of $Z$, the estimator obtained with \eqref{3.1.1} and \eqref{3.1.2} is not necessarily optimal. In spite of the loss of efficiency, there are two superior points to use the proposed estimating equations, that is, model identifiability and unbiasedness of $\theta$.
%\end{rem}

%\begin{rem}
%Instead of using  $h^* = E^{\star} \{ U( Z; \theta) \mid x; \phi_0 \}$ in (\ref{cc}), one can also consider using
%$$ \sum_{i=1}^n \left[ r_i U(z_i; \theta) + (1-r_i) E^{\star} \{ U( Z; \theta) \mid x_i; \phi_0 \} \right] = 0 $$
%as an optimal estimating equation for $\theta$. It can be shown that the two methods are asymptotically equivalent as long as $f( y \mid x)$ is correctly specified. (Is it correct statement?)(No, this is wrong. -KM)
%\end{rem}

\section{Adaptive Estimators}
\label{sec5}

We now propose two adaptive estimators for $(\phi_0, \theta_0)$: (i) with a parametric working model for $f_1(y \mid x)$; (ii) with a nonparametric estimator  for $f_1(y \mid x)$, where $f_1(y\mid x)=f(y\mid x, r=1)$. Although the optimality result in Lemma \ref{Lem.eff} has already been discussed by \citet{rotnitzky97}, the adaptive estimators proposed here are different from those of \citet{rotnitzky97}. See Appendix \ref{sup.C} for some  discussion of  \citet{rotnitzky97} estimator.

To discuss the first proposed method,
let $f_1(y\mid x)$ be known up to the parameter $\gamma\in\Gamma$, and let $\hat{\gamma}$ be the maximizer of $\sum_{i=1}^n r_i \log f_1(y_i\mid x_i; \gamma)$. This can be easily implemented, and the model selection can be implemented by using information criteria such as the Akaike information criterion (AIC) and the Bayesian information criterion (BIC). By using the idea similar to that used to derive \eqref{2.3}, we can show that, for any function $g(z)$,
\begin{align}
E^{\star}\{g(Z)\mid x; \phi_0,\gamma\} &= \frac{E_1\{\pi^{-1}(Z; \phi_0)O(Z; \phi_0) g(Z)\mid x; \gamma\}}{E_1\{\pi^{-1}(Z; \phi_0)O(Z; \phi_0)\mid x; \gamma\}}\label{4.1.1},
\end{align}
where $E_1(\cdot\mid x)=E(\cdot \mid x, r=1)$. Thus, the expectation can be estimated by using $f_1(y\mid x; \hat{\gamma})$ and $\pi(z; \phi_0)$. However, since $\phi_0$ is unknown, we propose an efficient estimating equation $\sum_{i=1}^n S_{\eff, i}(\phi, \theta,\hat{\gamma})=0$, where
\begin{align}
S_{\eff, i}(\phi, \theta,\hat{\gamma}) = \{S_{1}^\mathrm{T} (r_i, G_{r_i}(z_i); \phi,  \hat{\gamma}), S_{2}(r_i, G_{r_i}(z_i);\phi, \theta, \hat{\gamma})\}^\mathrm{T}, \label{eff.eq}
\end{align}
with
\begin{align*}
\begin{split}
S_{1}(r, G_r(z); \phi; \hat{\gamma})&=\llp 1-\frac{r}{\pi(z;\phi)}\rrp E^\star\{s_0(z;\phi)\mid x_i; \phi, \hat{\gamma}\},\\
S_{2}(r, G_r(z); \phi, \theta, \hat{\gamma})&=\frac{r}{\pi(z;\phi)} U(z;\theta)+\llp 1-\frac{r_i}{\pi(z;\phi)}\rrp E^{\star}\{U(z;\theta)\mid x_i; \phi, \hat{\gamma}\}.
\end{split}
\end{align*}
What if $f_1(y\mid x)$ is misspecified? One might expect the solution to the estimating equation with \eqref{eff.eq} to be inconsistent as a result. Note that the estimator that uses the function on the right-hand side of \eqref{4.1.1} is consistent even when the assumed model for $f_1(y\mid x)$ is misspecified. Also, if the model is correctly specified, the estimator attains the lower bound. This leads to Theorem \ref{thm.eff.bound}.

\begin{thm}
\label{thm.eff.bound}
Let $(\hat{\phi}^\mathrm{T}, \hat{\theta})^\mathrm{T}$ be the solution to  $\sum_{i=1}^n S_{\eff, i}(\phi, \theta,\hat{\gamma})=0$ in \eqref{eff.eq}. Under conditions \ref{I1}--\ref{I3} and  \ref{C1}--\ref{C7} given in Appendix \ref{sup.A}  and the identification conditions assumed in Theorem \ref{th.identification}, $(\hat{\phi}^\mathrm{T}, \hat{\theta})^\mathrm{T}$ satisfies consistency and asymptotic normality with variance
\begin{align*}
E\llp\frac{\pd S^*_\eff}{\pd (\phi^\mathrm{T}, \theta)}\rrp^{-1}E(S_\eff^{*\otimes 2})E\llp\frac{\pd S^*_\eff}{\pd (\phi^\mathrm{T}, \theta)^\mathrm{T}}\rrp^{-1},
\end{align*}
even if $f_1(y\mid x; \hat{\gamma})$ is misspecified, where $\gamma^*$ is the probability limit of $\hat{\gamma}$, and $S^*_\eff=\{S_1(\phi_0, \gamma^*)^\mathrm{T},$ $S_2(\phi_0, \theta_0, \gamma^*)\}^\mathrm{T}$ is defined in \eqref{eff.eq}.
In particular, the asymptotic variance of $\hat{\theta}$ is given as
\begin{align}
V^*=\mathrm{var}\left[\tau^{-1}_{\mathrm{U}}\{S_{2}(\phi_0, \theta_0, \gamma^*)-\kappa^*S_{1}( \phi_0, \gamma^*)\}\right], \label{vstar}
\end{align}
where $\kappa^*=\kappa^*_1(\kappa^*_2)^{-1}$, $\kappa^*_1=E[\{U^{\star}(\phi_0,\theta_0, \gamma^*)-U(\theta_0)\}\dot{\pi}(\phi_0)^\mathrm{T}/\pi(\phi_0)\}]$, \\$\kappa^*_2 = E\{g^{\star}(\phi_0, \gamma^*) \dot{\pi}(\phi_0)^\mathrm{T}/\pi(\phi_0)\}$, and $\tau_{\mathrm{U}}=E\{\pd U(\theta_0)/\pd \theta\}$. In addition, if the model is correctly specified, the estimator attains the semiparametric efficiency bound.
\end{thm}

Note that Theorem \ref{thm.eff.bound} does not require that $f_1$ be correctly specified. Unlike the estimator of \citet{riddles16}, the parametric model $f_1$ is irrelevant to the consistency and asymptotic normality of the estimator here. Therefore, we call $f_1$ a working model, as in \citet{liang86}. Also, though equation \eqref{3.1.2} has a form similar to that of the doubly robust estimator under MAR \citep{robins94}, our estimator does not have the doubly robustness  property. This is because the computation for $E^\star(\cdot\mid x)$ relies on the correct response mechanism.

%A modification is needed for $g^\star(x;\phi, \hat{\gamma})$ because it could be almost zero for some large values of $\phi$. Therefore, we recommend to multiply $g^\star_j(x;\phi, \hat{\gamma})$ times $\{\min_{i}g^\star_j(x_i;\phi, \hat{\gamma})\}^{-1}$ for each $j=1,\ldots, q$ so that $g^\star(x;\phi, \hat{\gamma})$ does not vanish for $\phi\neq \phi_0$. This modification does not effect on the efficiency because this gives exactly same estimating equation. It would be enough to conduct the modification only for  $j=1$ in practice.

Numerical computation is needed to calculate  the conditional expectation in (\ref{4.1.1}).  The expectation-maximization (EM) algorithm considered in  \citet{riddles16} can be used with a minor modification. We can directly apply their method, once the weights $w^*_{ij}$ defined in (15) of \citet{riddles16} are changed to
$$w^*_{ij}=\frac{r_j\pi^{-1}(x_{i},y_j;\phi)O(x_{i},y_j;\phi)f_1(y_j\mid x_i, \gamma)/C(y_j;\gamma)}{\sum_{k=1}^nr_k\pi^{-1}(x_{i},y_k;\phi)O(x_{i},y_k;\phi)f_1(y_k\mid x_i, \gamma)/C(y_k; \gamma)},$$
where $C(y;\gamma)=\sum_{l=1}^nr_l f_1(y\mid x_l;\gamma)$. The weight $w_{ij}^*$ can be called fractional weights in the context of fractional imputation of \citep{kim11}.  With these weights, $E^{\star}\{g(x_i, Y)\mid x_i;\gamma, \phi\}$ can be computed by $\sum_{j=1}^n w^*_{ij} g(x_i, y_j).$

We now discuss the second adaptive estimator based on nonparametric estimation for $f_1(y \mid x)$.  When $x$ is discrete, such as when $x$ is a binary variable, the expectation can be computed by averaging the data conditioned by $X=x$ and $R=1$, e.g., for $x=0, 1$,
\begin{align}
\hat{E}^{\star}\{g(x,Y)\mid x;\phi\} = \frac{\sum_{j\in I_x}r_j\pi^{-1}(x, y_j; \phi)O(x, y_j; \phi) g(x,y_j)}{\sum_{j\in I_x}r_j\pi^{-1}(x,y_j; \phi)O(x, y_j; \phi)} \label{ehat}
\end{align}
is a consistent estimator of \eqref{4.1.1}, where $I_x=\{j\in\{1,\ldots, n\}\mid X_j=x\}$.

When $x$ is continuous, the Nadaraya-Watson estimator can be employed. That is, for any function $g(z)$,
\begin{align}
\hat{E}^{\star}\{g(x, Y)\mid x; \phi\} &= \frac{\sum_{j=1}^n K_h (x-x_j)r_j \pi^{-1}(x, y_j; \phi)O(x, y_j; \phi) g(x, y_j)}{\sum_{j=1}^n K_h (x-x_j) r_j\pi^{-1}(x, y_j; \phi)O(x, y_j; \phi)}\label{4.1.3}
\end{align}
is consistent under the regularity conditions given in Appendix \ref{sup.A}. Here,
$K_h(x-w)=K\{(x-w)/h\}$, where $K$ is a kernel function, and $h$ is the bandwidth. We have the following result for the adaptive estimators obtained with the Nadaraya-Watson estimation.

\begin{thm}
\label{eff.nonpara}
Let $(\hat{\phi}^\mathrm{T}, \hat{\theta})^\mathrm{T}$ be the solution to $\sum_{i=1}^n \hat{S}_{\eff, i}(\phi, \theta)=0$, where $\hat{S}_{\eff, i}(\phi, \theta)$ is defined in \eqref{3.1.3} with the estimated conditional expectation in \eqref{4.1.3}.
Under Conditions \ref{I1}--\ref{I3}, \ref{C1}--\ref{C4}, and, \ref{C8}--\ref{C13} given in Appendix \ref{sup.A}, $(\hat{\phi}^\mathrm{T}, \hat{\theta})^\mathrm{T}$ satisfies consistency and asymptotic normality, and the estimator attains the semiparametric efficiency bound.
\end{thm}

\begin{rem}
The second proposed estimator is robust because it does not require any model assumptions on $f_1$, but it would not work well when the dimension of $x$ is high, as is common in any nonparametric estimation.
\end{rem}

%\begin{remark}
%When both $x$ and $y$ are discrete, the fully nonparametric approach proposed in Appendix 2 of \citet{riddles16} can be used. This approach makes it possible to estimate the response probability without specifying any parametric model. It can also be easily implemented with EM algorithm. In estimating response probability, the fully nonparametric approach is exactly the same as our proposed estimator. The proof is given in \S S3 in the Supplementary Material. Therefore, their  fully nonparametric approach is fully efficient and easier to implement.
%\end{remark}

%For example, if both $X$ and $Y$ are binary, $\pi(x,y)=P(R=1\mid x,y)$ takes four possible values. Since the fully nonparametric approach is based on the maximum likelihood, it is the same as the solution of \eqref{2.3} if the estimates of all $\pi(x,y)$ satisfy $0<\hat{\pi}(x,y)<1$; otherwise, there are no solutions. However, the estimator of the fully nonparametric approach has a solution if $\pi(x,y)>0$ while our proposed estimator does not have solutions unless the estimated probability is smaller than 1. In that sense, the fully nonparametric approach is better than our proposed method. Also, the fully nonparametric approach has exactly the same solution as our proposed estimator when the condition $0<\hat{\pi}(x,y)<1$ holds. The proof is given in \S S3 in the Supplementary Material. Therefore, when $X$ and $Y$ are discrete, the fully nonparametric approach always has a solution and it gives the best estimator.

Variance estimation is also a difficult problem in semiparametric estimation.  When we consider a parametric working model $f_1(y\mid x; \gamma)$,
\begin{align}
\hat{V}=n^{-1}\sum_{i=1}^n \left[ \hat{\tau}^{-1}_{\mathrm{U}}\{S_2(r_i, G_{r_i}(z_i); \hat{\phi},   \hat{\theta}, \hat{\gamma})-\hat{\kappa} S_1(r_i, G_{r_i}(z_i); \hat{\phi}, \hat{\gamma})\}\right]^{\otimes 2} \label{Vhat}
\end{align}
converges to $V^*$ in probability as defined in \eqref{vstar}, where $\hat{\tau}_{\mathrm{U}}$ and $\hat{\kappa}$ are consistent estimators for $\tau_{\mathrm{U}}$ and  $\kappa^*=\kappa^*_1(\kappa^*_2)^{-1}$, respectively, for $\kappa^*_1$ and $\kappa^*_2$ as defined in Theorem \ref{thm.eff.bound}. To estimate  $\kappa^*_1$, we propose using the same method that we used to compute $\theta_0$, i.e., let  $\mathcal{U}(\phi_0, k_1, \gamma^*)=k_1-(U^{\star}(\gamma^*)-U)\dot{\pi}(\phi_0)^\mathrm{T}/\pi(\phi_0)$ be our new $U$-function and let the solution to $E\{\mathcal{U}(\phi_0, k_1, \gamma^*)\} =0$ with respect to $k_1$ be our target parameter; solve the following equation:
$$\sum_{i=1}^n \lllp \frac{r_i}{\pi(z_i; \hat{\phi})}\mathcal{U}(z_i; \hat{\phi},k_1,\hat{\gamma})+\llp 1- \frac{r_i }{\pi(z_i; \hat{\phi})}\rrp E^{\star}\{\mathcal{U}( Z;  \hat{\phi}, k_1, \hat{\gamma})\mid x_i; \hat{\gamma}\}\rrrp =0.$$
This is the optimal estimator for $(\phi_0, \kappa^*_1)$ in terms of the asymptotic variance, because $\mathcal{U}$ is a known function and Theorem \ref{thm.eff.bound} is applicable. The best estimator for $\kappa^*_2$ can be obtained in the same way. When we use the nonparametric method stated in Theorem \ref{eff.nonpara} to estimate $\theta_0$, the variance can be also estimated by using the nonparametric method \eqref{ehat} and \eqref{4.1.3}, instead of using the parametric model $f_1(y\mid x;\gamma)$ in \eqref{Vhat}.

%%======Section6========%%
\section{Simulation Study}
\label{sec6}

\begin{figure}[h!]
	 \begin{center}
	\includegraphics[width=130mm]{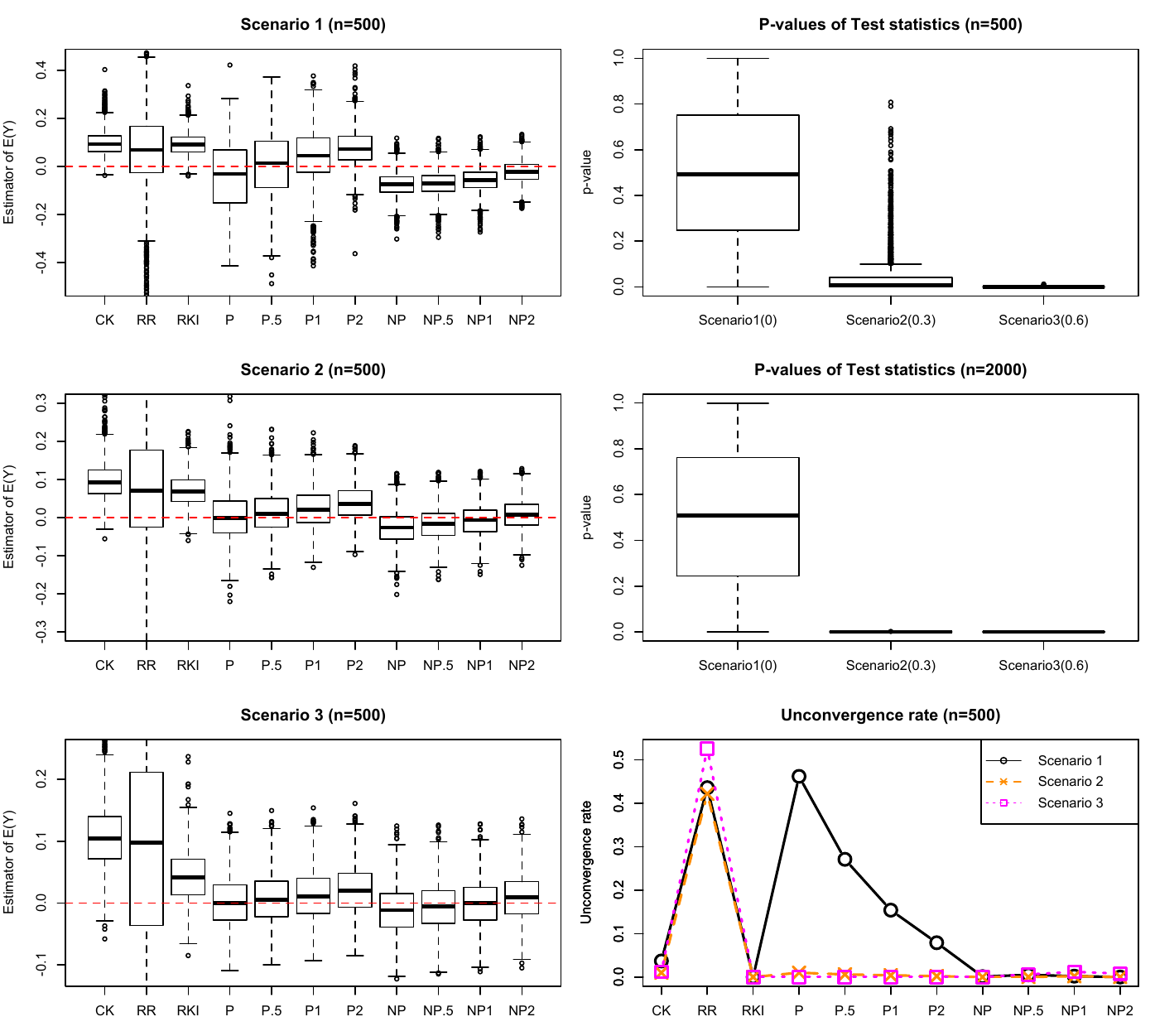}
	  \end{center}
	\caption{Left panel: Boxplot of Monte Carlo results for converged estimators for $\theta=E(Y)$ under three scenarios with sample size 500. The six estimators are CK (Chang \& Kott's estimator), RR (Rotnitzky \& Robins's estimator), RKI (Riddles' estimator), P (our proposed estimator with parametric $f_1$), NP (our proposed estimator with nonparametric $f_1$), and  P and NP method with DNET (doubly-normalized exponential transformation) where the suffix stands for $a$ value of $\mathcal{T}_a(x_1)$. The broken line shows the true value. Right panel: Test statistics for identifiability for datasets with each sample size 500 and 2000, and the bottom right panel shows the rate of estimators failed to obtain an esitmator.}
	\label{fig:2}
\end{figure}

{In order to evaluate the performance of our proposed estimators and to compare their efficiency with other methods in finite samples, we conduct a Monte Carlo simulation study with three scenarios. In each scenario, two covariates $X_1\sim N(0, 1/\sqrt{2}^2)$ and $X_2\mid X_1=x_1\sim N(-x_1/3, 1/2^2)$ are used.  For each scenario $s\;(=1,2,3)$, the response mechanism is  set to a Bernoulli distribution with parameter $\pi^{[s]}(x_1,x_2,y)$, where  $\pi^{[s]}(x_1, x_2, y)=1/\{1+\exp(\phi^{[s]}_{\mathrm{x}0}+0.5x_1+0.5x_2+\phi^{[s]}_{\mathrm{y}} y)\}$, and  the response outcome variables are generated from $Y\mid (x, r=1)\sim N(\mu^{[s]}(x), 1/2^2)$, where $\mu^{[s]}(x)=a^{[s]}_0+0.4x_1+0.4x_2+a_1^{[s]}x_1x_2$. The coefficients of the nonlinear term, which is the degree of nonlinearity, are set to $a_1^{[1]}=0, a_1^{[2]}=0.3, a_1^{[3]}=0.6$, and the other parameters are set, so that the expectation of the outcome variable is zero and the marginal response probability is 70\%, to $\phi^{[1]}_{\mathrm{x}0}=-0.959, \phi^{[2]}_{\mathrm{x}0}=-0.914, \phi^{[3]}_{\mathrm{x}0}=-0.904, \phi^{[1]}_{\mathrm{y}}=0.75,  \phi^{[2]}_{\mathrm{y}}=0.4,  \phi^{[3]}_{\mathrm{y}}=0.3, a_0^{[1]}=-0.0563, a_0^{[2]}=0.02$, and $a_0^{[3]}=0.0775$. Note that the scenario 2 and 3 are identifiable without using any instrumental variable because of the nonlinear term $x_1x_2$ in $f_1$, on the other hand, Scenario 1 is unidentifiable, and Scenario 2 is weakly identified than Scenario 3. We estimate $\theta= E(Y)$, thus $U(\theta; Z)=\theta-Y$, with two different Monte Carlo samples of size $n=500$ and $n=2000$ being independently generated 2,000 times. 

In Scenario 1, however, it is still possible to make the response model identifiable at the risk of misspecification of the response mechanism by using DNET. In this article, we change the variable $x_1\to\mathcal{T}_a(x_1)$ and $x_2\to\mathcal{T}_a(x_2)\;(a=0.5,1,2)$. 

From each sample, we compute six estimators, as follows:
\begin{enumerate}
\renewcommand{\labelenumi}{[\arabic{enumi}]}
%\item MAR: A naive estimator based on the assumption that the missing data are missing-at-random:
%\begin{align}
% \sum_{i=1}^n \delta_i(\theta-y_i)/\hat{\pi}_i=0, \label{HW}
%\end{align}
% where $\hat{\pi}_i$ is an estimated response mechanism, that is, $\hat{\pi}_i=\{1+\exp(\hat{\phi}_{\mathrm{x}0}+\hat{\phi}_{\mathrm{x}} x_i)\}^{-1}$, where $(\hat{\phi}_{\mathrm{x}0}, \hat{\phi}_{\mathrm{x}})$ is the maximum likelihood estimator.
\item CK: The estimator of \citet{chang08}. We use the estimating equation \eqref{ck}, setting $g$ as $(1, x_1, x_2)$; $\theta$ is estimated by solving
\begin{align}
 \sum_{i=1}^n r_i(\theta-y_i)/\hat{\pi}_i=0, \label{HW}
\end{align}
where $\hat{\pi}$ is the estimated response model.
\item {RR: The estimator of \citet{rotnitzky97}. This estimator is defined through four steps (i)--(iv) in Appendix \ref{sup.C}. In the first step, a consistent estimator is set to be the CK estimator, and in the second step, each of  \eqref{rr1}, \eqref{rr3}--\eqref{rr6} is modeled by at most third order polynomial function of $x_1$ and $x_2$. }
\item RKI: The estimator of \citet{riddles16}. In all scenarios, we specify a parametric model on $f_1$ based on normal distribution with the correct mean structure $\mu(x)=\beta_0+\beta_1x_1+\beta_2 x_2+\beta_3 x_1x_2. $
\item P: Our proposed estimator with parametric $f_1$ model. As  for the working model for $f_1$, the same model specification as in the RKI method is used. 
\item NP: Our proposed estimator with nonparametric $f_1$ model. As for the kernel function and its bandwidth, Gaussian kernel, and a rule-of-thumb bandwidth $h_j=n_1^{-1/5}\hat{\sigma}_{x_j}\:(j=1,2)$ is used, where $n_1$ is the sample size of observed outcome variable and $\hat{\sigma}_j$ is the square root of the sample variance of $x_j$ for $j=1,2$. 
\item DNET($a$): Same method as P and NP with the nonlinearly transformed data $\mathcal{T}_a(x_1)\;(a=0.5, 1, 2)$ for the variables in response models.
\end{enumerate}
\noindent
Suppose that the response model is correctly specified in our proposed methods, and except for our proposed methods, for identifiability, suppose that $x_2$ is specified as the instrumental variable, i.e., the response model is specified as
$$\mathrm{logit}\llp \pi(x_1,y)\rrp=\phi_{\mathrm{x}0}+\phi_{\mathrm{x}1}x_1+\phi_{\mathrm{y}} y.$$
%We used the correct models $\pi_{\mathrm{y}}(y)$ and $\pi_{\mathrm{xy}}(x,y)$ for the response mechanism, except for MAR.

%\begin{figure}
%	 \begin{center}
%	\includegraphics[width=105mm]{Fig2.pdf}
%	  \end{center}
%	\caption{Boxplot of Monte Carlo results for $\phi_\mathrm{y}$ and $\theta\{=E(Y)\}$ under four scenarios when $\phi_{\mathrm{x}1}$ is estimated. The four estimators are MAR (missing at random), CK (Chang \& Kott's estimator), RKI (Riddles' estimator), P (our proposed estimator with parametric $f_1$ model) NP (our proposed estimator with nonparametric method). Numbers 1 and 2 stand for $n=500$ and $n=2,000$, respectively. The broken line shows the true value.}
%\end{figure}

Before estimating the parameters, we first check the model identifiability of our proposed method. The right panel in Figure \ref{fig:2} shows the p-values of the statistical tests proposed in \S\ref{s3.2} under the three scenarios with different sample sizes. The p-values in scenario 1 spread around $1/2$ because the null hypothesis is correct or the model is unidentifiable. On the other hand, as the nonlinearity increases, p-values are close to zero. In particular, when $n=2000$, model identification can be judged with the probability almost 1 even for Scenario 2 which has a small degree of the nonlinearity $0.3$. 

The left panel in Figure \ref{fig:2} shows the Monte Carlo simulation results with sample size $n=500$. The results with sample size $n=2000$ are omitted because they are almost the same. In some Monte Carlo samples, we encounter some numerical problems and there is no solution because the estimate of the response model does not converge due to weak identifiability. The rates of datasets not having converging estimators are reported at the bottom right in Figure \ref{fig:2}.  The following is a summary of the simulation results:}

\begin{enumerate}
\renewcommand{\labelenumi}{[\arabic{enumi}]}
\item The CK method estimates the parameter stably, but it is biased due to the misspecification of the response model. The standard error of the CK estimators is a little larger than RKI, P, NP, and DNET methods due to the lack of efficiency.
\item {In many cases, the RR estimators do not converge. This comes from the difficulty of finding a good starting value of $\phi$ in the first step and of modeling the working models defined in Appendix \ref{sup.C}.}
\item Performance of RKI method is similar to that of CK, but it is less biased and has a smaller standard error because the $f_1$ model is correctly specified.
\item When the model is identifiable, the proposed P method works well. However, when it is unidentifiable, it is hard even to get a convergent sequence of estimators, though that can be inferred by testing linearity of the mean function.
\item Surprisingly, the NP method can estimate the parameter stably despite of the unidentifiability of the model, and the estimates are biased according to the degree of linearity of $\mu^{[s]}(x)$.
\item Proposed DNET works well for all the transformations $\mathcal{T}_{a}(x)$. In Scenario 1, when the model is not identifiable, the rate obtaining a non-convergent estimator and bias increase as the nonlinearity increases, in the meantime, the standard error decreases.
\end{enumerate}

\section{Real data analysis}
\label{sec7}

In this section, our proposed estimators are applied to the Korea Labor and Income Panel Survey (KLIPS) data, which have been analyzed multiple times \citep{kim11_2, wang14, shao16}. The data contain $n=2,506$ Korean wage earners; the response variable $y$ is total wage income ($10^6$ Korean Won) in year 2008. There are three fully observed covariates: $x_1$: total wage income in the previous year (2007); $x_2$: gender; $x_3$: age. While $x_1$ is a continuous variable, $x_2$ has two categories 1 and 2 for male and female, respectively, and $x_3$ has three categories 1-3: $x_3<35,\;35\leq x_3 <51$, and $\;x_3\geq 51$. We also identified three data points as outliers and excluded them from further analysis.

Although the data are completely observed, we took the approach of \citet{kim11_2} and created 1000 incomplete datasets with the following eight response mechanisms: M1 (linear nonignorable without $(x_2,x_3)$): $\logit(\pi)=0.48-0.3 x_1-0.5y$; M2 (linear nonignorable): $\logit(\pi)=-0.85-0.2 x_1+0.5 x_2+0.2 x_3 -0.4 y$; M3 (nonlinear nonignorable, quadratic in $x_1$ without $(x_2,x_3)$): $\logit(\pi)=0.33-0.3 x_1-0.1x_1^2-0.3 y$; M4 (nonlinear nonignorable, quadratic in $x_1$): $\logit(\pi)=-0.89-0.4 x_1-0.1x_1^2+0.5 x_2+0.2 x_3-0.4 y$;  M5 (nonlinear nonignorable, quadratic in $y$ without $(x_2,x_3)$): $\logit(\pi)=0.24-0.25 x_1-0.25 y-0.1y^2$; M6 (nonlinear nonignorable, quadratic in $y$): $\logit(\pi)=-0.93-0.2 x_1+0.5 x_2+0.2 x_3-0.2 y-0.1y^2$;  M7 (probit nonignorable)  $\pi=\Phi(-0.55+0.3x_1+0.4y)$;  M8 (jump nonignorable) $\pi=0.5 I(0.5 x_2+0.2 x_3+y\leq 2.6)+0.9 (0.5 x_2+0.2 x_3+y>2.6)$, where $\Phi(\cdot)$ is the cumulative distribution function of the standard normal distribution, and $I(A)$ is the indicator function that takes 1(0) if event $A$ is true (false). Note that there are NIVs for models M2, M4, M6, and M8. For all data sets, the response rate is about 70\%. We estimated $\theta=E(Y)$ as considered in the simulation. The ``true" average income in 2008 is $\hat{\theta}_{n}=1.846$  as calculated using the complete data. In order to estimate the parameters, we assumed a response mechanism
$\logit\{\pi(x, y; \phi)\}=\phi_{\mathrm{x}0}+\phi_{\mathrm{x}1} x_1+\phi_{\mathrm{x}2} x_2+\phi_{\mathrm{x}3} x_3+\phi_{\mathrm{y}} y$. Therefore M1 and M2 are correctly specified while M3-M8 are misspecified. %Created eight response indicators $R_{j}\;(j=1, \ldots, 8)$ are available from ``https://github.com/KosukeMorikawa/NMAR/blob/master/r\_dat.csv".

We specified unknown $f_1$ models as normal distribution $Y\mid (x_1, x_2=i, x_3=j, r=1) \sim N(\mu_{i,j}(x_1), \sigma^2_{i,j})$ ($i=1,2$; $j=1,2,3$), where
$\mu_{i,j}(x_1)=\gamma_{0i,j}+\gamma_{1i,j}x_{1}+\gamma_{2i,j}x_{1}^2+\gamma_{3i,j}x_{1}^3+\gamma_{4i,j} x_{1}^4$; $(\gamma_{1i,j}, \gamma_{2i,j}, \gamma_{3i,j}, \gamma_{4i,j})$ is the regression parameter when $(x_2, x_3)=(i,j)$. We chose the best model by AIC among $2^5-1$ models for each $(x_2, x_3)$'s $2\times 3$ pattern. Using Theorem \ref{th.identification}, one can show that this model is identifiable as one of the 6 mean structures are nonlinear, or all the structures are linear but all of them are not the same. One simple sufficient condition is to check whether the conditional mean of $y$ given $x_1$ is linear with respect to $x_1$. In the real data, the correlation between $x_1$ and $y$ is too high because wage income does not change considerably within one year; the mean structure is almost linear. However, the p-values of the test statistics are almost zero in all datasets with M1--M8, therefore, all the response models are identifiable without using any instrumental variable nor transformation. In Table \ref{tab:2}, Bias, S.E. (standard error), and RMSE (root mean square error) with five methods, CK, RR, RKI, P, NP methods same as in \S 5, are reported. Following are summary of the results:

\begin{enumerate}
\renewcommand{\labelenumi}{[\arabic{enumi}]}
\item The CK method estimates the parameter stably, but it is inefficient compared to our proposed methods.
\item As in \S 5, the RR estimators do not converge in many datasets. 
\item RKI methods can obtain estimates stably, but it is severely biased due to the misspecification of $f_1$ model, which is generally unknown in real data. 
\item The proposed P method works well, but for some datasets, we encounter some numerical problems due to the misspecification of the response model. As for such datasets, we may get a reasonable estimator by using DNET. Note that our method is effective for the probit response mechanism (M7), even though the use of the probit model makes it hard to identify the parameter as stated in \S 3.1.
\item Performance of the proposed NP method is the best among the five methods considered. However, the results with dataset M5 and M6 implies the difficulty of obtaining the estimator with misspecified response models.
\end{enumerate}
%Therefore, to obtain valid estimator of $\theta$, we considered two different approaches: [1] find NIVs used; [2] transform $x_1$ with DNET in the response model. For the first approach, we  specified $x_2$, $x_3$, and $(x_2,x_3)$ as instrumental variables in applying our proposed method, which will lead to inconsistency for models M2 because there is actually no instrumental variable. For the second approach, we transformed $x_1$ to $\mathcal{T}_1(x_1)$ so that \eqref{id.cond} becomes identifiable even if the cumulant-generating function is normally distributed with a linear mean structure. Although this transformation made the model identifiable, this also changed the assumed mechanism to $\logit\{\pi(x, y; \phi)\}=\phi_{\mathrm{x}0}+\phi_{\mathrm{x}1} \mathcal{T}_1(x_1)+\phi_{\mathrm{x}2} x_2+\phi_{\mathrm{x}3} x_3+\phi_{\mathrm{y}} y$. This may be a potential cause of biased estimation. On the flip side, this approach uses all information of covariates, which helps to reduce bias and gain efficiency. We show the result of this approach under both parametric and nonparametric $f_1$ models.

\begin{table}
\begin{center}
\caption{Bias, S.E. (standard error), and RMSE (root mean square error) of our proposed estimator, where the full sample estimate $\hat{\theta}_{n}=1.846$ is set to the true value, for datasets M1--M8. NA rate is the rate of dataset failed to get a convergence estimator. All values are multiplied by 1,000 except for NA rate.}
\begin{tabular}{clrrrrr} \hline \hline
\label{tab:2}
\multirow{2}{*}{Model} & & \multicolumn{5}{c}{Methods}  \\ \cline{3-7} 
 && CK & RR & RKI & P & NP\\ \hline
\multirow{4}{*}{M1}  & Bias            & 73 & 118 & 737 & 22 & 17\\
                            & S.E.            & 130 & 388 & 676 & 48 & 31\\
                            & RMSE         & 149 & 406 & 1000 & 52 & 35\\
                            & NA rate(\%)& 0.7 & 31.9 & 0 & 1.8 & 1.6\\ \hline
\multirow{4}{*}{M2}  & Bias            & 66 & 90 & 599 & 23 & 13\\
                            & S.E.            & 119 & 217 & 580 & 49 & 31\\
                            & RMSE         & 136 & 235 & 834 & 54 & 33\\
                            & NA rate(\%)& 1.1 & 27.9 & 0 & 1.2 & 1.0\\ \hline
\multirow{4}{*}{M3}  & Bias            & 205 & 197 & 879 & 11 & -6\\
                            & S.E.            & 128 & 694 & 810 & 49 & 31\\
                            & RMSE         & 241 & 722 & 1195 & 50 & 32\\
                            & NA rate(\%)& 8.1 & 48.4 & 0 & 4.3 & 2.5\\ \hline
\multirow{4}{*}{M4}  & Bias            & -103 & 10 & 245 & 55 & 29\\
                            & S.E.            & 100 & 179 & 295 & 51 & 26\\
                            & RMSE         & 144 & 179 & 383 & 75 & 39\\
                            & NA rate(\%)& 15.8 & 40.9 & 0 & 1.1 & 0.3\\ \hline
\multirow{4}{*}{M5}  & Bias            & -39 & 78 & 840 & 59 & 0\\
                            & S.E.            & 71 & 288 & 780 & 211 & 56\\
                            & RMSE         & 81 & 298 & 1146 & 219 & 56\\
                            & NA rate(\%)& 2.9 & 24.2 & 0 & 1.9 & 32.7\\ \hline
\multirow{4}{*}{M6}  & Bias            & -68 & 57 & 776 & 44 & 13\\
                            & S.E.            & 87 & 550 & 734 & 51 & 49\\
                            & RMSE         & 111 & 553 & 1068 & 68 & 51\\
                            & NA rate(\%)& 7.4 & 34.0 & 0 & 1.7 & 37.4\\ \hline
\multirow{4}{*}{M7}  & Bias            & 158 & 125 & 1131 & 15 & 11\\
                            & S.E.            & 155 & 472 & 844 & 42 & 34\\
                            & RMSE         & 221 & 489 & 1412 & 45 & 36\\
                            & NA rate(\%)& 2.2 & 36.7 & 0 & 5.8 & 2.0\\ \hline
\multirow{4}{*}{M8}  & Bias            & 175 & 136 & 689 & 27 & 9\\
                            & S.E.            & 115 & 475 & 579 & 48 & 33\\
                            & RMSE         & 209 & 494 & 900 & 55 & 34\\
                            & NA rate(\%)& 0.6 & 37.2 & 0 & 1.4 & 1.1\\ \hline \hline
\end{tabular}
\end{center}
\end{table}

%\begin{table}
%\begin{center}
%\caption{$\hat{\theta}-\hat{\theta}_{n}$ (S.E. $(\hat{\theta})$): deviation of our proposed estimator $\hat{\theta}$ from the full sample estimate $\hat{\theta}_{n}=1.846$ (and estimated their standard error) for datasets M1--M8 by two approaches: [1] using instrumental variable (IV); [2] using transformed $\mathcal{T}_1(x_1)$ with parametric (P) and nonparametric (NP)  $f_1$ model. NA stands for not applicable due to numerical problems. All values are multiplied by 1,000.}
%\begin{tabular}{ccccccc} \hline
%\label{tab:2}
%& \multicolumn{6}{c}{Approach}   \\
%& \multicolumn{3}{c}{[1] NIV}  & & \multicolumn{2}{c}{[2] DNET}   \\ \cline{2-4}  \cline{6-7}
%IV &  $x_2$ & $x_3$ & $(x_2, x_3)$ && &    \\
% method&  P & P & P && P & NP   \\ \hline
%M1 & \,  -8 (48) &    \ 16 (33)        & \, NA (NA) &           &  -13 (24)          &\, -9 (34)\\
%M2 & \;\;  -73 (1118) & \, -8 (37)        & \;  -54 (115) &          & \, -1 (24)          & \;\,  5 (26)\\
%M3 & -25 (33) & \ NA (NA)      & \,  46 (48) &       & \, -8 (24)      & \,\; 6 (25)\\
%M4 & \;\,-19 (255) & \  13 (90)        & \;\;\; -97 (2654) &           & \ 18 (27)             & \ 20 (27)\\
%M5 &  \ 26 (49) & \  32 (89)     &\, 60 (34) &     & -11 (23)     & -27 (48)\\
%M6 &  \ 50 (82) & \ 183 (147)       &  161 (72) &        &\, -7 (23)        & -14 (21)\\
%M7 & \ 41 (55) &\ 56 (47)   &\,  55 (42) && \ 26 (30)  & \ 15 (47)\\
%M8 & \ 23 (65) &  -10 (29)    & \,   NA (NA) & & \,  -1 (26)   & \,\; 2 (23)\\
%\hline
%\end{tabular}
%\end{center}
%\end{table}

%%======Section7========%%
\section{Discussion}
We have presented a test statistic for model identification, semiparametric efficiency bound  for $(\phi^\mathrm{T}_0, \theta_0)^\mathrm{T}$ under nonignorable nonresponse; proposed two types of adaptive semiparametric estimators that attain the semiparametric lower bound. Identification is a challenging problem in nonignorable nonresponse \citep{miao16}; previous methods require nonignorable NIVs to guarantee model identification \citep{wang14}. Our new identifiability condition is not on the response mechanism, but on the distribution of $[y\mid x, r=1]$. %The identifiability of the assumed response model can be found by checking the distribution without using any NIV.

The proposed method is based on the correct specification of the response model. There may be various other models for the true response mechanism, and thus the appropriate information criteria for choosing the response mechanism will be a topic of future research. Instead of specifying a single response model,  one can consider multiple response models, and obtain consistency  when one of the specified response models is correct.  This multiple robustness property has been investigated under the ignorable nonresponse setup \citep{han14,chen17}. Extension of multiple robustness to the nonignorable nonresponse case will also be a topic of our future research.

%\section*{Acknowledgements}
%The authors are grateful for the very constructive comments of the three anonymous referees and the Associate Editor.

\appendix

\section{Regularity conditions}
\label{sup.A}
\begin{enumerate}
\renewcommand{\theenumi}{(C\arabic{enumi})}
\item \label{C1}  $\Phi$ and $\Theta$ are compact.
\item \label{C2} $W_i=( X_i,Y_i, R_i)$ are independently and identically distributed.
\item \label{C3} $\Gamma$ is compact, $S_{\gamma}(\gamma)=\pd\log f_1(y\mid x;\gamma)/\pd \gamma$ is continuously differentiable at $\gamma\in\Gamma$ with probability one, there exists $e(W)$ such that $\|S_{\gamma}(\gamma)\|\leq e(W)$ for all $\gamma\in\Gamma$ and $E\{e(W)\}<\infty$, $E\{S_{\gamma}(\gamma)\}=0$ has a unique solution $\gamma^* \in \Gamma$, $\pd S_{\gamma}(\gamma)/\pd \gamma^\mathrm{T}$ is continuous at $\gamma^*$ with probability one, and there is a neighborhood $\Gamma_\mathcal{N}$ of $\gamma^*$ such that $\|E\{\sup_{\gamma\in\Gamma_{\mathcal{N}}}\pd S_{\gamma}(\gamma)/\pd\gamma^\mathrm{T}\}\|<\infty$.
\item \label{C4} Identifiability of $\theta$  for complete data: there exists  $\theta_0\in\Theta$ such that $E\{U(Z; \theta_0)\}=0$.
%\item \label{C5} $E\{S_\eff(\phi, \theta,\gamma^*)\}=0$ has a unique solution $(\phi_0,\theta_0)\in\Phi\times\Theta$, where $S_\eff(\phi,\theta,\gamma)=(S_1(\phi,\gamma)^\mathrm{T}, S_2(\phi, \theta,\gamma))^\mathrm{T}$ defined in (10).
\item \label{C5} $ \pd S_\eff(\phi, \theta,\gamma)/\pd(\phi^{\mathrm{T}},\theta, \gamma^\mathrm{T})$ is continuous at $(\phi_0, \theta_0, \gamma^*)$ with probability one, and there is a neighborhood $\Phi_{\mathcal{N}}\times \Theta_{\mathcal{N}}\times \Gamma_{\mathcal{N}}$ of $(\phi_0, \theta_0,\gamma^*)$ such that $$\|E\{\sup_{(\phi,\theta, \gamma^*)\in\Phi_{\mathcal{N}}\times\Theta_{\mathcal{N}}\times\Gamma_{\mathcal{N}}} \pd S_\eff(\phi, \theta, \gamma)/\pd(\phi^{\mathrm{T}},\theta,\gamma^\mathrm{T})\}\|<\infty.$$
\item \label{C6} $S_\eff(\phi, \theta,\gamma)$  is continuously differentiable at each $(\phi, \theta, \gamma)\in\Phi\times\Theta\times\Gamma$ with probability one, and there exists $d_1(W)$ such that $\|S_\eff(\phi, \theta, \gamma)\|\leq d_1(W)$ for all $(\phi, \theta,\gamma)\in\Phi\times\Theta\times\Gamma$ and $E\{d_1(W)\}<\infty$.
%\item \textcolor{red}{$E[\{1-E(\pi(Z; \phi_0)/\pi(Z; \phi)\mid X)\} g^\star(X;\phi)] \neq 0$ for all $\phi\neq \phi_0$} (I am not sure whether we can assume this condition, which is needed to assure uniqueness of our estimator).
\item \label{C7} $E\{ \pd S_\eff(\phi, \theta, \gamma^*)/\pd(\phi^{\mathrm{T}},\theta, \gamma^\mathrm{T})\}$ is nonsingular at $(\phi_0, \theta_0, \gamma^*)$.
\item \label{C8} The conditions (C5)-(C7) hold at the true value $\gamma^*=\gamma_0$.
\item \label{C9} Let $\mathcal{X}$ be the support of $x$. Then,  $f_1(x)>0$ and  $E_1\{\pi(x,Y;\phi_0)\mid x\}>0$ for all $x\in\mathcal{X}$.
\item \label{C10} The kernel $K(u)$ has bounded derivatives of order $k$, satisfies $\int K(u)\mathrm{d}u = 1$, has zero moments of order $\leq m-1$, and has a nonzero $m$-th order moment.
\item \label{C11} For all $y$, $\pi(\cdot, y;\phi_0)$, $\dot{\pi}(\cdot, y; \phi_0)$, and $U(\cdot, y; \theta_0)$ are differentiable to order $k$ and are bounded on an open set containing $\mathcal{X}$.
\item \label{C12} Let $a_1(z)=1$, $a_2(z)=s_0(z;\phi_0)$, and $a_3(z)=U(z)$. Then, there exists $v\geq 4$ such that $E_1\{|\pi^{-1}(Z;\phi_0)O(Z;\phi_0)a_i(Z)|^v\}$ and $E_1\{\|\pi^{-1}(Z;\phi_0)O(Z;\phi_0)a_i(Z)\|^v\mid x\}f_1(x)$ are bounded for all $x\in \mathcal{X}$.
\item \label{C13} As $h\to 0$, $n^{1-(2/v)}h^d/\ln n\to \infty, n^{1/2}h^{d+2k}/\ln n\to \infty$, and $n^{1/2}h^{2m}\to0$.
\end{enumerate}

%%======Appendix B========%%
\section{Proofs of the technical results}
\label{sup.B}

\noindent
\textit{Proof of Theorem \ref{th.identification}.}   Let $f_1(y\mid x)$ be the true density function of $[y\mid x, r=1]$. Here, the distribution of $[y\mid x]$ can be represented through the observed outcome density and the response model, because by using Bayes' formula, we have
\begin{align}
f(y\mid x; \phi)=\frac{f_1(y\mid x) \pi^{-1}(x, y; \phi)}{\int f_1(y\mid x) \pi^{-1}(x, y; \phi)dy}.\label{f_f1}
\end{align}
Suppose that $\phi_0$ is the true value of the response model so that the true distribution of $[y\mid x]$ is $f(y\mid x; \phi_0)$. Then, it follows from \eqref{f_f1} that the probability limit of the estimating equation is 
\begin{align*}
	E\llp \Gamma(Z, R;\phi)\mid x\rrp
	&= g(x; \phi)\int \llp 1-\frac{\pi(Z; \phi_0)}{\pi(Z; \phi)}\rrp f(y\mid x; \phi_0) dy\\
	&= g(x; \phi) \llp 1-\frac{\int \pi(Z; \phi)^{-1}f(y\mid x; \phi) dy}{\int \pi(Z; \phi_0)^{-1}f(y\mid x; \phi_0) dy}\rrp.
\end{align*}
By using \ref{I2} and \ref{I3}, the conditional expectation can not be vanished unless $\phi=\phi_0$. Therefore, the solution is unique.
\qed
\vspace{1ex}

\noindent
\textit{Proof of Proposition \ref{prop3.1}.}
For any error function $\varepsilon\in\mathcal{E}$, under the null hypothesis $H^{(\infty)}_0$, there exist $c^{(\ell)}_1, c^{(\ell)}_2\; (\ell=2,3)$ such that $E(\varepsilon^2\mid x)=c^{(2)}_1+(c^{(2)}_2)^\top x$ and $E(\varepsilon^3\mid x)=c^{(3)}_1+(c^{(3)}_2)^\top x$. On the other hand, it holds that
\begin{align*}
\varepsilon^2=\sum_{j=0}^\infty \xi_j^2 e_j^2(x)+\sum_{j\neq k}\xi_j\xi_k e_j(x)e_k(x).
\end{align*}
It follows from $e_j\neq e_k\;(j\neq k)$  that there must exist a positive integer $j$ such that $e_j=\{E(\xi_j^2)\}^{-1/2}(c^{(2)}_1+(c^{(2)}_2)^\top x)^{1/2}$ and $e_k\equiv 0$ for $k\neq j$, so that $\varepsilon=(c^{(2)}_1+(c^{(2)}_2)^\top x)^{1/2}\xi_j$. Let such $j$ be 1 without loss of generality. In a similar way, it follows from the third moment condition of $\varepsilon$ that $\varepsilon=(c^{(3)}_1+(c^{(3)}_2)^\top x)^{1/3}\xi_1$, which implies $c^{(2)}_2=c^{(3)}_2=0$ and $c^{(3)}_1=(c^{(2)}_1)^{3/2}$. By using the induction, under the null hypothesis $H^{(\infty)}_0$, it can be shown that $c^{(\ell)}_2\equiv 0$ for $\ell\geq 2$.  As a result, $\varepsilon$ is a random variable which is independent of  $x$. Therefore, $H^{(1)}_0$ or testing linearity of mean function is enough to check the model identification. \qed
\vspace{1ex}

Next, we provide a proof of Lemma \ref{Lem.eff} and Theorem \ref{thm.eff.bound} and \ref{eff.nonpara}.  In order to prove Lemma \ref{Lem.eff}, we will assume $U(z)=y$ just for simplicity. We specify the joint distribution $z=(x^\mathrm{T}, y)^\mathrm{T}$ by $f(z; \eta)$, where $\eta$ is an infinite-dimensional nuisance parameter, and $\eta_0$ is the true value. By ``full model" we refer to the class of models in which the data are completely observed, and by ``obs model" we refer to those in which some $Y$ are missing; that is, a full model consists of functions $h(Z)$ and an obs model consists of $h(R, G_R(Z))$. Furthermore, for each full and obs model, denote the nuisance tangent space by $\Lambda^F$ and $\Lambda$, respectively, and its orthogonal complement by $\Lambda^{F\perp}$ and $\Lambda^\perp$, respectively. Let $S_\phi$ be the score function with respect to $\phi$. Consider a Hilbert space $\mathcal{H}=\{h^{(q+1)\times 1}\mid E(h)=0; \|h\|<\infty \}$ with inner product $\langle h_1, h_2\rangle=E(h_1^\mathrm{T} h_2)$, where the expectation is taken under the true model.
See \citet{bickel98} and \citet{tsiatis06} for more details. When $U$ is comprised of other functions, the proof is almost the same.

At first,  we introduce a proposition of \citet{rotnitzky97}, which provides the efficient score for $(\phi, \theta)$, as follows. Let $B$ and $D$ be functions of $(R, G_{R}(Z))$, and let $B^*$ and $D^*$ be functions of $Z$. Also, let us define the following three linear operators: $\rmg(B^*)=E(B^*\mid R, G_{R}(Z))$, $\rmm(B^*)=E\{\rmg(B^*)\mid Z\}$, and $\rmu(B^*)=RB^*/\pi(Z)$. Then, the efficient score for $(\phi, \theta)$ can be derived by the following Lemma. See Proposition A1 in \citet{rotnitzky97} for the proof.

\begin{lem}
\label{sup.lem1}
The efficient score for $(\phi, \theta)$ can be written as
\begin{align}
 S_\eff=\rmu(D^*_\eff)-\Pi[\rmu(D^*_\eff)\mid \Lambda_2]+A_{2,\eff}=\rmg\{\rmm^{-1}(D^*_\eff)\}+A_{2,\eff}, \label{B1}
\end{align}
where $\Pi[h\mid \Lambda_2]$ is the projection of $h$ onto $\Lambda_2$, $\Lambda_2=[h(R,G_R(Z)): E(h(R,G_R(Z))\mid Z)=0]$, and $D^*_\eff$ is a unique solution to
\begin{align}
\Pi[\rmm^{-1}(D^*)\mid \Lambda^{F\perp}]=(Q, S^{F\perp}_{\eff,\theta}), \label{B2}
\end{align}
where $Q=\Pi[\rmm^{-1}[E\{\rmg(S^F_{\phi})\mid L\}]\mid \Lambda^{F\perp}]$, $A_{2,\eff}=(\Pi[S_{\phi}\mid \Lambda_2]^\mathrm{T}, 0)^\mathrm{T}=(\rmg(S^F_\phi)-\rmg[\rmm^{-1}[E\{\rmg(S^F_\phi)\mid L\}]]^\mathrm{T}, 0)^\mathrm{T}$, and $S^{F\perp}_{\eff,\theta}$ is the efficient score function of $\theta$ in the full model.
\end{lem}

This Lemma implies that the efficient score can be represented by \eqref{B1} with $D^*_\eff$ satisfying condition \eqref{B2}.
Thus, in the nonignorable nonresponse case, $\Lambda^{F\perp}$ needs to be calculated, and it can be done in a way similar to that shown in Section 4.5 of \citet{tsiatis06}.

\begin{lem}
\label{sup.lem2}
The nuisance tangent space $\Lambda^F$ and its orthogonal complement $\Lambda^{F\perp}$ in the full model are written as follows:
\begin{align*}
\Lambda^F&=[h(Z)\in\mathcal{H}~ \mathrm{such~that~} E\{Yh(Z)\}=0], \\
 \Lambda^{F\perp}&=\lllp k(Y-\theta_0), \mathrm{where}~k\mathrm{~is~any~}q+1\mathrm{~dimensional~vector}\rrrp.
\end{align*}
\end{lem}

Finally, we give an explicit formula to calculate the projection onto $\Lambda_2$.

\begin{lem}
\label{sup.lem3}
For $h(R,G_R(Z))=Rh_1(Z)+(1-R)h_2(X)$, it holds that
\begin{align}
\Pi(h\mid \Lambda_2)=\llp 1-\frac{R}{\pi(Z)}\rrp \frac{E\lllp \{1-\pi(Z)\}\{h_2(X)-E\{h_1(Z)\} \mid X\rrrp}{E\{O(Z)\mid X\}}. \label{B3}
\end{align}
\end{lem}

\noindent
\textit{Proof of Lemma \ref{sup.lem3}.}
Obviously, the right-hand side of \eqref{B3} belongs to $\Lambda_2$. Thus, it remains to check that for any $g$,
$$\left\langle h-\llp1-\frac{R}{\pi(Z)}\rrp \frac{h_2(X)-E\{h_1(Z)\mid X\}}{E\{O(Z)\mid X\}}, \llp1-\frac{R}{\pi(Z)}\rrp g(X)\right\rangle=0,$$
which can be proved easily. \qed

\noindent
We now give a proof of Lemma \ref{Lem.eff}.

\vspace{1ex}

\noindent
\textit{Proof of Lemma \ref{Lem.eff}.}
 Note that $S^{F\perp}_{\eff, \theta}=Y-\theta_0$ by Lemma \ref{sup.lem2}, since there exists only one influence function, and it is the efficient one under the assumption that $\theta$ does not require any assumptions on the distribution of $Z$ \citep[see][Chap. 5]{tsiatis06}. By the projection theorem, there exists a unique $k=(k_1, k^\mathrm{T}_2)^\mathrm{T}$ such that $D^*_\eff=k (Y-\theta_0)$.

Then, we calculate $A_{2,\eff}$. The score function of $\phi$ is
$$S_{\phi}=\rmg(S^F_{\phi})=Rs_1(Z; \phi)+(1-R)s_0(X;\phi),$$
where $s_r(\phi)$ is defined in (3). It follows from  Lemma \ref{sup.lem3} with $h_1(z)=s_1(\phi)$ and $h_2(x)=\bar{s}_0(x;\phi)$ in \eqref{B3} that $\Pi(S_{\phi}\mid \Lambda_2)=-\{1-R/\pi(Z)\}g^{\star}(X)$.
Thus, $A_{2,\eff}=[0, -\{1-R/\pi(Z)\}g^{\star}(X)].$
Again, by using Lemma \ref{sup.lem3}, it follows that $\Pi[\rmu(D^*_\eff)\mid \Lambda_2]=-\{1-R/\pi(Z)\}E^{\star}(Y-\theta_0\mid X)$, by which \eqref{B1} becomes
\begin{align*}
S_1=k_2\lllp\frac{R(Y-\theta_0)}{\pi(\phi_0)}+\llp 1- \frac{R}{\pi(\phi_0)}\rrp E^{\star}(Y-\theta_0\mid X)\rrrp-\llp 1- \frac{R}{\pi(Z)}\rrp g^{\star}(X)
\end{align*}
and
\begin{align*}
S_2=k_1\lllp\frac{R(Y-\theta_0)}{\pi(\phi_0)}+\llp 1- \frac{R}{\pi(\phi_0)}\rrp E^{\star}(Y-\theta_0\mid X)\rrrp.
\end{align*}
This $S_\eff=(S_1, S^\mathrm{T}_2)$ can be transformed into $\tilde{S}_\eff=(\tilde{S}_1, \tilde{S}^\mathrm{T}_2)=AS_\eff$,
\begin{align*}
\tilde{S}_1 &= \llp 1- \frac{R}{\pi(\phi_0)}\rrp g^{\star}(X),\\
\tilde{S}_2 &= \frac{R(Y-\theta_0)}{\pi(\phi_0)}+\llp 1- \frac{R}{\pi(\phi_0)}\rrp E^{\star}(Y-\theta_0\mid X)
\end{align*}
with a nonsingular matrix $A$,
\begin{align*}
A=\begin{bmatrix}
-I_q & -k^\mathrm{T}_2/k_1 \\
0^\mathrm{T} & k_1^{-1} \\
\end{bmatrix},
\end{align*}
where $I_q$ is a $q$-dimensional identity matrix. The score function multiplied by a nonsingular constant matrix does not have an influence on the asymptotic distribution. Thus, we have the desired efficient score. \qed
\vspace{1ex}

\noindent
\textit{Proof of Theorem \ref{thm.eff.bound}.} Consistency and asymptotic normality are proved under the assumptions \ref{C1}--\ref{C8} by using the standard argument for GMM. Next, we give the explicit form of the asymptotic variance.
Let $\xi=(\phi^\mathrm{T}, \theta)^\mathrm{T}$. Recall that each $\hat{\gamma}$ and $\hat{\xi}$ is a solution to $\sum_{i=1}^n \pd \log f_1(y_i\mid x_i;\gamma)/\pd \gamma=\sum_{i=1}^n S_{\gamma i}(\gamma)=0$ and $\sum_{i=1}^n S_{\eff, i}(\hat{\gamma},\xi)=0$, respectively, where $S_{\eff, i}(\gamma,\xi)$ is defined in (10). By using standard asymptotic theory,
$$\begin{bmatrix}
\hat{\gamma}-\gamma^* \\
\hat{\xi}-\xi_0
\end{bmatrix}=-\mathcal{I}^{-1} n^{-1}\sum_{i=1}^n \begin{bmatrix}
 S_{\gamma i}(\gamma^*) \\
S_{\eff,i}(\gamma^*, \xi_0)
\end{bmatrix},
$$
where
\begin{align*}
\mathcal{I}&=E
\begin{bmatrix}
\pd S_{\gamma}(\gamma^*)/ \gamma^\mathrm{T} & \pd S_{\gamma}(\gamma^*)/ \xi^\mathrm{T} \\
\pd S_\eff(\gamma^*, \xi_0)/ \gamma^\mathrm{T} & \pd S_\eff(\gamma^*, \xi_0)/ \xi^\mathrm{T}
\end{bmatrix}\\
&=E
\begin{bmatrix}
\pd S_{\gamma}(\gamma^*)/ \gamma^\mathrm{T} & {O} \\
\pd S_\eff(\gamma^*, \xi_0)/ \gamma^\mathrm{T} & \pd S_\eff(\gamma^*, \xi_0)/ \xi^\mathrm{T}
\end{bmatrix}.
\end{align*}
Let the $(i,j)$ block of $\mathcal{I}$ be $\mathcal{I}_{ij}$. Then,
$$
\mathcal{I}^{-1}=\begin{bmatrix}
\mathcal{I}^{-1}_{11} & {O}\\
-\mathcal{I}^{-1}_{22}\mathcal{I}_{21}\mathcal{I}^{-1}_{2} & \mathcal{I}^{-1}_{22}
\end{bmatrix}.
$$
Here, it follows that $\mathcal{I}_{21}={O}$ because
$$E\lllp \llp 1- \frac{R}{\pi(\phi_0)}\rrp \frac{\pd g^{\star}(\gamma^*, \xi_0)}{\pd \gamma^\mathrm{T}}\rrrp={O}$$
and
$$E\lllp \llp 1- \frac{R}{\pi(\phi_0)}\rrp \frac{\pd U^{\star}(\gamma^*,  \xi_0)}{\pd \gamma^\mathrm{T}}\rrrp=0^\mathrm{T}.$$
Therefore, we have,
$$
\mathcal{I}^{-1}
=\begin{bmatrix}
\mathcal{I}^{-1}_{11} & O\\
O & \mathcal{I}^{-1}_{22}
\end{bmatrix}.
$$
By applying exactly the same arguments for $\mI^{-1}_{22}$ used for $\mI^{-1}$, we got the asymptotic variance of $\hat{\theta}$ as given in (11).
\qed
\vspace{1ex}
%\textit{Proof of Proposition 1.}
%It follows from the definition of $g^{\star}$ and $f_1$ that
%\begin{align}
%g^{\star\mathrm{T}}(x)&=\frac{E_1\{O(\phi; x, Y)(\dot{h}(x; \phi_{\mathrm{x}})^\mathrm{T},\; Y)\mid x\}}{E_1\{\pi^{-1}(\phi;x,Y)O(\phi; x, Y)\mid x\}}\nonumber\\
%&=\frac{( \dot{h}(x;\phi_{\mathrm{x}})^\mathrm{T} E_1\{\exp(\phi_{\mathrm{y}} Y)\mid x\} ,\;E_1\{Y\exp(\phi_{\mathrm{y}} Y)\})}{E_1[\{1+\exp(h(x; \phi_{\mathrm{x}})+\phi_{\mathrm{y}} Y)\}\exp(\phi_{\mathrm{y}} Y)\mid x]}, \label{pp1}
%\end{align}
%where $\dot{h}(\phi_{\mathrm{x}})=\pd h(\phi_{\mathrm{x}})/\pd \phi_{\mathrm{x}}$. Thus, $E_1\{\exp(\phi_{\mathrm{y}} Y)\mid x\}$ and $E_1\{Y\exp(\phi_{\mathrm{y}} Y)\mid x\}$ are to be computed. Since $E_1\{\exp(\phi_{\mathrm{y}} Y)\mid x\}$ is the generating moment function of $Y\mid x$, it holds that
%$$E_1(\exp(\phi_{\mathrm{y}} Y)\mid x)=\exp\left(\frac{b(\phi_{\mathrm{y}}\psi+\theta)-b(\theta)}{\psi}\right)$$
%and
%$$E_1(Y\exp(\phi_{\mathrm{y}} Y)\mid x)=\dot{b}(\phi_{\mathrm{y}}\psi+\theta)\exp\left(\frac{b(\phi_{\mathrm{y}}\psi+\theta)-b(\theta)}{\psi}\right).$$
%Substituting these results into \eqref{pp1}, we obtain the proof. Computation of $E^{\star}(Y\mid x)$ can be done in a similar way. \qed

\noindent
\textit{Proof of Theorem \ref{eff.nonpara}.}
Consistency and asymptotic normality of our proposed estimator are similar to proving Lemma \ref{Lem.eff} of \citet{morikawa17}. We herein show our estimator attains the semiparametric lower bound derived in Lemma \ref{Lem.eff}. Let $f_1(x)$ be the conditional distribution of $[x\mid r=1]$. From the same arguments that were used to prove Lemma A.1 in \citet{morikawa17}, it can be shown  that the estimating equation in Theorem \ref{eff.nonpara}, $\hat{S}_{\eff}(\phi,\theta)=\{\hat{S}_1(\phi)^\mathrm{T}, \hat{S}_2(\phi,\theta)\}^\mathrm{T}$ is expanded as
\begin{align*}
\hat{S}_1(\phi)&=n^{-1}\sum_{i=1}^n\lllp\llp1-\frac{r_i}{\pi(\phi;z_i)}\rrp g^{\star}(\phi;x_i)+r_iG(z_i; \phi)\rrrp+o_p(n^{-1/2})\\
\hat{S}_2(\phi,\theta)&=n^{-1}\sum_{i=1}^n\lllp \frac{r_i}{\pi(\phi;z_i)}U(\theta; z_i)+\llp1-\frac{r_i}{\pi(\phi;z_i)}\rrp U^\star(\theta, \phi;x_i)+r_i H(\theta, \phi; z_i)\rrrp\\
&\quad +o_p(n^{-1/2}),
\end{align*}
where $G(\phi; z_i)=G_1(\phi; x_i)G_2(\phi; z_i)$,  $H(\theta, \phi; z_i)=G_1(\phi; x_i)H_2(\theta, \phi; z_i)$, and
\begin{align*}
G_1(\phi; x_i)&= 1-E\llp\frac{\pi(\phi_0; Z)}{\pi(\phi; Z)} \;\bigg|\; x_i\rrp,\\
G_2(\phi; z_i)&=\frac{\pi^{-1}(\phi;z_i)O(\phi;z_i)\{s_0(\phi;z_i)-g^{\star}(\phi;x_i)\}}{E_1\{\pi^{-1}(\phi; Z)O(\phi; Z)\mid x_i\}P(R=1\mid x_i)},\\
H_2(\theta, \phi; z_i)&=\frac{\pi^{-1}(\phi;z_i)O(\phi;z_i)\{U(\theta; z_i)-U^\star(\theta, \phi; x_i)\}}{E_1\{\pi^{-1}(\phi; Z)O(\phi; Z)\mid x_i\}P(R=1\mid x_i)}.
\end{align*}

Therefore, the asymptotic variance may increase due to the additional terms $rG(\phi)$ and $rH(\phi)$, but this solution also attains the lower bound. At first, we focus on the estimator for $\phi$. Once we get an unbiased estimating equation $\sum_{i=1}^n \varphi(z_i;\phi)=0$, the asymptotic variance can be given as $\mathrm{Var}\{E(\dot{\varphi}(\phi_0))^{-1}\varphi(\phi_0)\}$, where $\dot{\varphi}(\phi_0)=\pd {\varphi}(\phi_0)/ \pd \phi^\mathrm{T}$. Thus, for the proving purpose, it suffices to show that $G(\phi_0)=0$ and  $E(R \dot{G}(\phi_0))={O}$. The former equation is trivial, so we only need to work on the latter equation, which can be written as $E(R \dot{G}(\phi_0))=E(R G_1(\phi_0)\dot{G}_2(\phi_0))+E(R G_2(\phi_0)\dot{G}_1(\phi_0))$.  The first term is zero from $G_1(\phi_0)=0$. Also, the second term is $E(R G_2(\phi_0)\dot{G}_1(\phi_0)) =E\{E(R G_2(\phi_0)\mid X)\dot{G}_1(\phi_0)\}={O}$. Hence, the last equation holds by the definition of $g^\star(\phi; x)$. Therefore, $rG(\phi)$ has no effect on the asymptotic variance and our estimator also attains the semiparametric efficiency bound. The same conclusion can be made when estimating $\theta$.
\qed

%%======Appendix C========%%
\section{Comparison with Rotnitzky and Robins (1997)'s estimator}
\label{sup.C}

{
In \citet{rotnitzky97}, the semiparametric efficiency bound for NMAR data was derived in more general settings in Proposition A1 and A2, and  an adaptive estimator for regression coefficients was proposed. However, to attain the efficiency bound,  the estimator requires many working models to be correctly specified, and it would be pratically impossible to correctly specify all of the models. For example, for the case of nonignorable nornesponse, seven working models, equations (32) to (38) in \citet{rotnitzky97},  have to be specified.}

 In particular, if $\theta=E(Y)$ is our parameter of interest,  three working models are required:
\begin{align}
E_1\{\pi^{-1}(Z; \phi_0)O(Z; \phi_0)\mid x\} &=: \nu_1(x; \zeta_1),\label{rr1}\\
E_1\{\pi^{-1}(Z; \phi_0)O(Z; \phi_0)s_0(Z; \phi_0)\mid x\} &=: \nu_2(x; \zeta_2),\label{rr2}\\
E_1\{Y\pi^{-1}(Z; \phi_0)O(Z; \phi_0)\mid x\} & =: \nu_3(x; \zeta_3).\label{rr3}
\end{align}
Note that \eqref{rr2} is a multi-dimensional function. For example, in the same setup as \S 6, i.e., $\mathrm{logit}\{\pi(x,y;\phi)\}=\phi_{\mathrm{x0}}+\phi_{\mathrm{x1}}x_1+\phi_{\mathrm{y}}y$, where $x=(x_1, x_2)$, \eqref{rr2} can be written as
\begin{align}
E_1\{O(Z; \phi_0)\mid x\} &=: \nu_4(x; \zeta_4),\label{rr4}\\
E_1\{x_1O(Z; \phi_0)\mid x\} &=: \nu_5(x; \zeta_5),\label{rr5}\\
E_1\{YO(Z; \phi_0)\mid x\} &=: \nu_5(x; \zeta_6),\label{rr6}
\end{align}
where $\nu_2(x; \zeta_2)=\{\nu_4(x; \zeta_4), \nu_5(x; \zeta_5), \nu_6(x; \zeta_6)\}^\top$.

Then an adaptive estimator of $\phi$ and $\theta$ can be obtained from the  following four steps:
\begin{enumerate}
\renewcommand{\theenumi}{(\roman{enumi})}
	\item Find a consistent estimator $\tilde{\phi}$ of $\phi_0$ by e.g. \citet{chang08}'s method;
	\item Estimate ${\zeta}_k\;(k=1, 2, 3)$ in (C.1)-(C.3) by the least square method with the estimated $\tilde{\phi}$;
	\item Let $\hat{\phi}$ be a solution to
	\begin{align*}
	\sum_{i=1}^n\llp 1-\frac{r_i}{\pi(z_i; \phi)}\rrp \frac{\nu_2(x_i; \hat{\zeta}_2)}{\nu_1(x_i; \hat{\zeta}_1)}=0.
	\end{align*}
	\item Let $\hat{\theta}$ be the solution to
	\begin{align*}
	\sum_{i=1}^n\lllp \frac{r_i(y_i-\theta)}{\pi(z_i; \hat{\phi})} +\llp 1-\frac{r_i}{\pi(z_i; \hat{\phi})}\rrp \llp \frac{\nu_3(x_i; \hat{\zeta}_3)}{\nu_1(x_i; \hat{\zeta}_1)}-\theta\rrp\rrrp=0.
	\end{align*}
\end{enumerate}

Therefore, their adaptive estimator is similar to the two-step estimator in GMM. However, as shown in section 6, it may be practically difficult to find a valid consistent estimator of $\phi$ for NMAR data. Also,  giving reasonable parametric models for \eqref{rr1}-\eqref{rr3} are challenging because the left-hand side of them are non-linear functions.


\begin{thebibliography}{}

\bibitem[{Berrett \& Samworth(2019)}]{berrett19}
\textsc{Berrett, T.~B..} \& \textsc{Samworth, R.~J.} (2019).
\newblock {Nonparametric independence testing via mutual information}
\newblock \textit{Biometrika} \textbf{106}, 547--571.
\MR{3992389}

\bibitem[{Bickel et~al.(1998)}]{bickel98}
\textsc{Bickel, P. J.}, \textsc{Klaassen, C.~A.~J.}, \textsc{Ritov, Y.}, \& \textsc{Wellner, J.~A.} (1998).
\newblock \textit{{Efficient and Adaptive Estimation for Semiparametric Models}}.
\newblock New York: Springer-Verlag.
\MR{1623559}

\bibitem[{Chang \& Kott(2008)}]{chang08}
\textsc{Chang, T.} \& \textsc{Kott, P. S.} (2008).
\newblock {Using calibration weighting to adjust for nonresponse under a plausible model}
\newblock \textit{Biometrika} \textbf{95}, 555--571.
\MR{2443175}

\bibitem[{Chen \& Haziza(2017)}]{chen17}
\textsc{Chen, S.} \& \textsc{Haziza, D.} (2017).
\newblock {Multiply robust imputation procedures for the treatment of item nonresponse in surveys.}
\newblock \textit{Biometrika} \textbf{104}, 439--453.
\MR{3698264}

\bibitem[{D'Haultfoeuille(2010)}]{dhau10}
\textsc{D'Haultfoeuille, X.} (2010).
\newblock {A new instrumental method for dealing with endogeneous selection}.
\newblock \textit{J. Econometrics} \textbf{154}, 1--15.
\MR{2558947}

\bibitem[{Diggle \& Kenward(1994)}]{diggle94}
\textsc{Diggle, P.} \& \textsc{Kenward, M.~G.}  (1994).
\newblock {Informative drop-out in longitudinal data analysis}.
\newblock \textit{J. R. Statist. Soc. {\rm C}} \textbf{43}, 49--93.

\bibitem[{Dom\'{i}nguez \& Lobato(2004)}]{dominguez04}
\textsc{Dom\'{i}nguez}, M.~A. \& \textsc{Lobato, I.~N.}  (2004).
\newblock {Consistent estimation of models defined by conditional moment restrictions}.
\newblock \textit{Econometrica} \textbf{72}, 1601--1615.
\MR{2078215}

\bibitem[{Eubank \& Hart(1992)}]{eubank92}
\textsc{Eubank, R. L.} \& \textsc{Hart, J. D.}  (1992).
\newblock {Testing Goodness-of-fit in regression via order selection criteria}.
\newblock \textit{Ann. Statist.} \textbf{20}, 1412--1425.
\MR{1186256}

\bibitem[{Fitzmaurice et~al.(2005)}]{fitzmaurice05}
\textsc{Fitzmaurice, G.~M.}, \textsc{Lipsitz, S.~R.}, \textsc{Molenberghs, G.} \& \textsc{Ibrahim, J.~G.} (2005).
\newblock {A protective estimator for longitudinal binary data subject to non-ignorable non-monotone missingness}.
\newblock \textit{J. R. Statist. Soc. {\rm A}} \textbf{168}, 723--735.
\MR{2205403}

\bibitem[{Fukumizu et~al.(2004)}]{fukumizu04}
\textsc{Fukumizu, K.}, \textsc{Bach, F.~R.}, and \textsc{Jordan, M.~I.} (2004).
\newblock {Dimensionality reduction for supervised learning with reproducing kernel Hilbert spaces}.
\newblock \textit{J. Mach. Learn. Res.} \textbf{5}, 73--99.
\MR{2247974}

\bibitem[{Greenlees et~al.(1982)}]{greenlees82}
\textsc{Greenlees, J.~S.}, \textsc{Reece, W.~S.} \& \textsc{Zieschang, K.~D.} (1982).
\newblock {Imputation of missing values when the probability of response depends on the variable being imputed}.
\newblock \textit{J. Am. Statist. Assoc.} \textbf{77}, 251--261.

\bibitem[{Gretton et~al.(2005)}]{gretton05}
\textsc{Gretton, A.}, \textsc{Herbrich, R.}, \textsc{Smola, A.}, \textsc{Bousque, O.} \& \textsc{Sch\"olkopf, B.} (2005).
\newblock {Kernel methods for measuring independence}.
\newblock \textit{J. Mach. Learn. Res.} \textbf{6}, 2075--2129.
\MR{2249882}

\bibitem[{Gretton et~al.(2008)}]{gretton08}
\textsc{Gretton, A.}, \textsc{Fukumizu, K.}, \textsc{Teo, C.~H.}, \textsc{Song, L.}, \textsc{Sch\"olkopf, B.} \&  \textsc{Smola, A.}(2008).
\newblock {A kernel statistical test of independence}.
\newblock \textit{NeurIPS} \textbf{20}, 585--592.

\bibitem[{H\'ajek(1970)}]{hajek70}
\textsc{H\'ajek, J.} (1970).
\newblock {A characterization of limiting distributions of regular estimates}.
\newblock \textit{Z. Wahrscheinlichkeitstheorie verw. Gebiete.}  \textbf{14}, 323--330.
\MR{0283911}

\bibitem[{Han(2014)}]{han14}
\textsc{Han, P.} (2014).
\newblock Multiply robust estimation in regression analysis with missing data.
\newblock \textit{J. Am. Statist. Assoc.} \textbf{109}, 1159--1173.
\MR{3265688}

\bibitem[{Hidalgo et~al.(2018)}]{hidalgo18}
\textsc{Hidalgo, S.~J.~T.}, \textsc{Wu, M.~C.}, \textsc{Engel, S.~M.} \& \textsc{Kosorok, M.~R.}  (2018).
\newblock Goodness-of-fit test for nonparametric regression models: Smoothing spline ANOVA models as example.
\newblock \textit{Comput. Stat. Data An.} \textbf{122}, 135--155.
\MR{3765820}

\bibitem[{Hart(1997)}]{hart97}
\textsc{Hart, J~D.} (1997).
\newblock \textit{{Nonparametric Smoothing and Lack-of-fit tests}}.
\newblock New York: Springer-Verlag.
\MR{1461272}

\bibitem[{Kim(2011)}]{kim11}
\textsc{Kim, J.~K.} (2011).
\newblock {Parametric fractional imputation for missing data analysis}
\newblock \textit{Biometrika} \textbf{98}, 119--132.
\MR{2804214}

\bibitem[{Kim \& Yu(2011)}]{kim11_2}
\textsc{Kim, J.~K.} \& \textsc{Yu, C.~L.} (2011).
\newblock A semiparametric estimation of mean functionals with nonignorable missing data.
\newblock \textit{J. Am. Statist. Assoc.} \textbf{106}, 157--165.
\MR{2816710}

\bibitem[{Liang \& Zeger(1986)}]{liang86}
\textsc{Liang, K.-Y.} \& \textsc{Zeger, S.~L.} (1986).
\newblock {Longitudinal data analysis using generalized linear models}
\newblock \textit{Biometrika} \textbf{73}, 13--22.
\MR{0836430}

\bibitem[{Little \& Rubin(2002)}]{little02}
\textsc{Little, R.~J.~A.} \& \textsc{Rubin, D.~B.} (2002).
\newblock \textit{{Statistical Inference with Missing Data. Second edition}}.
\newblock New York: Wiley.
\MR{1925014}

\bibitem[{Louis(1982)}]{louis82}
\textsc{Louis, T.~A.} (1982).
\newblock {Finding the observed information matrix when using the EM algorithm}.
\newblock \textit{J. R. Statist. Soc. {\rm B}} \textbf{44}, 226--233.
\MR{0676213}

\bibitem[{Ma et~al.(2003)}]{ma03}
\textsc{Ma, W.~Q.}, \textsc{Geng, Z.} \& \textsc{Hu, Y.~H.} (2003).
\newblock {Identification of graphical models for nonignorable nonresponse of binary outcomes in longitudinal studies}.
\newblock \textit{J. Multivar. Anal.} \textbf{87}, 24--45.
\MR{2007260}

\bibitem[Miao et~al.(2016)]{miao16}
\textsc{Miao, W.}, \textsc{Ding, P.} \& \textsc{Geng, Z.} (2016).
\newblock Identifiability of normal and normal mixture models with nonignorable missing data.
\newblock \textit{J. Am. Statist. Assoc.} \textbf{111}, 1673--1683.
\MR{3601726}

\bibitem[{Molenberghs et~al.(2008)}]{molenberghs08}
\textsc{Molenberghs, G.}, \textsc{Beunckens, C.}, \textsc{Sotto, C.} \& \textsc{Kenward, M.~G.} (2008).
\newblock {Every missingness not at random model has a missingness at random counterpart with equal fit}.
\newblock \textit{J. R. Statist. Soc. {\rm B}} \textbf{70}, 371--388.
\MR{2424758}

\bibitem[{Morikawa et~al.(2017)}]{morikawa17}
\textsc{Morikawa, K.}, \textsc{Kim, J.~K.} \& \textsc{Kano, Y.} (2017).
\newblock {Semiparametric maximum likelihood estimation with data missing not at random}.
\newblock \textit{Canad. J. Statist.} \textbf{45}, 393--409.
\MR{3729977}

\bibitem[Qin et~al.(2002)]{qin02}
\textsc{Qin, J.}, \textsc{Leung, D.} \& \textsc{Shao, J.} (2002).
\newblock Estimation with survey data under nonignorable nonresponse or informative sampling.
\newblock \textit{J. Am. Statist. Assoc.} \textbf{97}, 193--200.
\MR{1947279}

\bibitem[Riddles et~al.(2016)]{riddles16}
\textsc{Riddles, M.~K.}, \textsc{Kim, J.~K.} \& \textsc{Im, J.} (2016).
\newblock Propensity-score-adjustment method for nonignorable nonresponse.
\newblock \textit{J. Surv. Stat. Methodol.} \textbf{97}, 215--245.

\bibitem[Robins et~al.(1994)]{robins94}
\textsc{Robins, J.~M.}, \textsc{Rotnitzky, A.} \& \textsc{Zhao, L.~P.} (1994).
\newblock Estimation of regression coefficients when some regressors are not always observed.
\newblock \textit{J. Am. Statist. Assoc.} \textbf{89}, 846--866.
\MR{1294730}

\bibitem[Robins et~al.(2000)]{robins99}
\textsc{Robins, J.~M.}, \textsc{Rotnitzky, A.} \& \textsc{Scharfstein, D.~O.} (2000).
\newblock \textit{Sensitivity Analysis for Selection Bias and Unmeasured Confounding in Missing Data and Causal Inference Models}.
\newblock \textit{In Statistical Models in Epidemiology: The Environment and Clinical Trials.}
\newblock New York: Springer-Verlag, 1--92.
\MR{1731681}

\bibitem[{Rotnitzky \& Robins(1997)}]{rotnitzky97}
\textsc{Rotnitzky, A.} \& \textsc{Robins, J.~M.} (1997).
\newblock {Analysis of semi-parametric regression models with non-ignorable non-response}.
\newblock \textit{Stat. Med.} \textbf{16}, 81--102.

\bibitem[{Rotnitzky et~al.(2001)}]{rotnitzky01}
\textsc{Rotnitzky, A.}, \textsc{Scharfstein, D.}, \textsc{Su, T.-L.} \& \textsc{Robins, J.~M.} (2001).
\newblock {Methods for conducting sensitivity analysis of trials with potentially nonignorable competing causes of censoring}.
\newblock \textit{Biometrics} \textbf{57}, 103--113.
\MR{1833295}

\bibitem[{Rubin(1976)}]{rubin76}
\textsc{Rubin, D.~B.} (1976).
\newblock {Inference and missing data}.
\newblock \textit{Biometrika} \textbf{61}, 581--592.
\MR{0455196}

\bibitem[Scharfstein et~al.(1999)]{scharfstein99}
\textsc{Scharfstein, D.~O.}, \textsc{Rotnitzky, A.} \& \textsc{Robins, J.~M.} (1999).
\newblock Adjusting for nonignorable drop-out using semiparametric nonresponse models.
\newblock \textit{J. Am. Statist. Ass.} \textbf{94}, 1096--1146.
\MR{1731478}

\bibitem[{Sen \& Sen(2014)}]{sen14}
\textsc{Sen, A.} \& \textsc{Sen, B.} (2014).
\newblock {Testing independence and goodness-of-fit in linear models}.
\newblock \textit{Biometrika} \textbf{101}, 927--942.
\MR{3286926}


\bibitem[{Shao \& Wang(2016)}]{shao16}
\textsc{Shao, J.} \& \textsc{Wang, L.} (2016).
\newblock {Semiparametric inverse propensity weighting for nonignorable missing data}.
\newblock \textit{Biometrika} \textbf{103}, 175--187.
\MR{3465829}

\bibitem[{Skrondal \& Rabe-Hesketh(2014)}]{skrondal14}
\textsc{Skrondal, A.} \& \textsc{Rabe-Hesketh, S.} (2014).
\newblock {Protective estimation of mixed-effects logistic regression when data are not missing at random}.
\newblock \textit{Biometrika} \textbf{101}, 175--188.
\MR{3180664}

\bibitem[{Sverchkov(2008)}]{sverchkov08}
\textsc{Sverchkov, M.}  (2014).
\newblock {A new approach to estimation of response probabilities when missing data are not missing at random}.
\newblock \textit{In Proc. Survey Res. Meth. Sect., Am. Statist. Ass.}
\newblock Washington DC: American Statistical Association, 867--874.

\bibitem[{Tang et~al.(2003)}]{tang03}
\textsc{Tang, G.}, \textsc{Little, R.~J.~A.} \& \textsc{Raghunathan, T.~E.} (2003).
\newblock {Analysis of multivariate missing data with nonignorable nonresponse}.
\newblock \textit{Biometrika} \textbf{90}, 747--764.
\MR{2024755}

\bibitem[{Tang et~al.(2014)}]{tang14}
\textsc{Tang, N.}, \textsc{Zhao, P.} \& \textsc{Zhu, H.} (2014).
\newblock {Empirical likelihood for estimating equations with nonignorably missing data}.
\newblock \textit{Statist. Sinica} \textbf{24}, 723--747.
\MR{3235396}

\bibitem[{Tsiatis(2006)}]{tsiatis06}
\textsc{Tsiatis, A.~A.}  (2006).
\newblock \textit{{Semiparametric Theory and Missing Data}}.
\newblock New York: Springer-Verlag.
\MR{2233926}

\bibitem[{Verbeke et~al.(2001)}]{verbeke01}
\textsc{Verbeke, G.}, \textsc{Molenberghs, G.}, \textsc{Thijs, H.}, \textsc{Lesaffre, E.} \& \textsc{Kenward, M.~G.} (2002).
\newblock {Sensitivity analysis for nonrandom dropout: A local influence approach}.
\newblock \textit{Biometrics} \textbf{57}, 7--14.
\MR{1833286}

\bibitem[{Wang et~al.(2014)}]{wang14}
\textsc{Wang, S.}, \textsc{Shao, J.} \& \textsc{Kim, J.~K.} (2014).
\newblock {An instrumental variable approach for identification and estimation with nonignorable nonresponse}.
\newblock \textit{Statist. Sinica} \textbf{24}, 1097--1116.
\MR{3241279}

\bibitem[{Yang et~al.(2017)}]{yang17}
\textsc{Yang, S.}, \textsc{Wang, L.} \& \textsc{Ding, P.} (2017).
\newblock {Nonparametric identification of causal effects with confounders subject to instrumental missingness}.
\newblock \textit{arXiv:1702.03951v1}.

\bibitem[{Zhao \& Shao(2015)}]{zhao15}
\textsc{Zhao, J.} \& \textsc{Shao, J.} (2015).
\newblock {Semiparametric pseudo-likelihoods in generalized linear models with nonignorable missing data}.
\newblock \textit{J. Am. Statist. Ass.} \textbf{110}, 1577--1590.
\MR{3449056}

\bibitem[{Zhao et~al.(2017)}]{zhao17}
\textsc{Zhao, P.}, \textsc{Tang, N.}, \textsc{Qu, A.} \& \textsc{Jiang, D.} (2017).
\newblock {Semiparametric estimating equations inference with nonignorable missing data}.
\newblock \textit{Statist. Sinica} \textbf{27}, 89--113.
\MR{3618161}
\end{thebibliography}
\end{document}